\documentclass[reqno,12pt,letterpaper]{amsart}
\pdfoutput=1
\usepackage{color, xcolor, colortbl}
\usepackage{graphicx,epstopdf}
\usepackage{amsmath,amssymb,amsthm}
\usepackage{algorithm}
\usepackage{algorithmic}
\usepackage{bm}
\usepackage[caption=false]{subfig}
\usepackage{appendix}
\usepackage{multirow}
\usepackage{braket}
\usepackage[english]{babel}
\usepackage{hyperref}
\usepackage[capitalize]{cleveref}
\usepackage[square,numbers,sort&compress]{natbib}
\usepackage{adjustbox}
\usepackage{xspace}
\usepackage[roman]{complexity}
\usepackage{qcircuit}

\newcommand\blfootnote[1]{%
  \begingroup
  \renewcommand\thefootnote{}\footnote{#1}%
  \addtocounter{footnote}{-1}%
  \endgroup
}

\newcommand{\eps}{\varepsilon}

\newcommand{\diag}{\operatorname{diag}}

\DeclareMathOperator*{\cinf}{C^{\infty}}
\DeclareMathOperator*{\esssup}{ess\,sup}
\newcommand{\dd}{\partial}
\newcommand{\la}{\langle}
\newcommand{\ra}{\rangle}

\newcommand{\I}{\mathrm{i}}

\newcommand{\wt}[1]{\widetilde{#1}}

\newcommand{\norm}[1]{\left\lVert#1\right\rVert}

\newcommand{\op}{{\rm op}}

\newcommand{\Or}{\mathcal{O}}

\newcommand{\NN}{\mathbb{N}}
\newcommand{\RR}{\mathbb{R}}

\newcommand{\ZZ}{\mathbb{Z}}
\newcommand{\TT}{\mathbb{T}}
\newcommand{\Sc}{\mathcal{S}}

\newtheorem{theorem}{\protect\theoremname}
\theoremstyle{plain}
\newtheorem{lemma}[theorem]{\protect\lemmaname}
\theoremstyle{plain}
\newtheorem{remark}[theorem]{\protect\remarkname}
\theoremstyle{plain}
\newtheorem*{lem*}{\protect\lemmaname}
\theoremstyle{plain}

\theoremstyle{plain}
\newtheorem{corollary}[theorem]{\protect\corollaryname}

\providecommand{\definitionname}{Definition}
\providecommand{\assumptionname}{Assumption}
\providecommand{\corollaryname}{Corollary}
\providecommand{\lemmaname}{Lemma}
\providecommand{\propositionname}{Proposition}
\providecommand{\remarkname}{Remark}
\providecommand{\theoremname}{Theorem}

\newcommand{\REV}[1]{\textcolor{black}{ #1}}
\newcommand{\REVV}[1]{\textcolor{black}{ #1}}

\title[Uniform observable error bounds of Trotter formulae]{Uniform observable error bounds of Trotter formulae for the semiclassical Schr\"odinger equation}

\author{Yonah Borns-Weil} \address{Department of Mathematics, University of California, Berkeley,  CA 94720, USA} \email{yonah\_borns-weil@berkeley.edu}

\author{Di Fang} \address{Department of Mathematics and Duke Quantum Center, Duke University, Durham, NC 27710, USA} \email{di.fang@duke.edu}
  
 \date{}

\begin{document}
\maketitle

\begin{abstract}
\REV{Known as} no fast-forwarding theorem \REV{in quantum computing},
the simulation time for the Hamiltonian evolution needs to be $\Or(\norm{H} t)$ in the worst case, which essentially states that one can not go across the multiple scales as the simulation time for the Hamiltonian evolution needs to be strictly greater than the physical time. We demonstrated in the context of the semiclassical Schr\"odinger equation that the computational cost for a class of observables can be much lower than the state-of-the-art bounds. In the semiclassical regime (the effective Planck constant $h \ll 1$), the operator norm of the Hamiltonian is $\Or(h^{-1})$. We show that the number of Trotter steps used for the observable evolution can be $\Or(1)$, that is, to simulate some observables of the Schr\"odinger equation on a quantum scale only takes the simulation time comparable to the classical scale. In terms of error analysis, we improve the additive observable error bounds [Lasser-Lubich 2020] to uniform-in-$h$ observable error bounds. This is, to our knowledge, the first uniform observable error bound for semiclassical Schr\"odinger equation without sacrificing the convergence order of the numerical method. Based on semiclassical calculus and discrete microlocal analysis, our result showcases the potential improvements taking advantage of multiscale properties, such as the smallness of the effective Planck constant, of the underlying dynamics and sheds light on going across the scale for quantum dynamics simulation.
\end{abstract}

\tableofcontents

\section{Introduction}

Multiscale problems are ubiquitous in nature and typically considered to be difficult tasks in scientific computation and numerical analysis due to the separation of scales and hence the existence of certain small parameters (see, e.g., \cite{E2011}).
One prominent example in quantum dynamics is the semiclassical Schr\"odinger equation, defined as
\begin{equation} \label{eq:semi_schd}
    \I h \partial_t u^h(t,x) = -\frac{h^2}{2} \Delta u^h(t,x) + V(x) u^h(t,x)  =: H u^h(t,x) ,
\end{equation}
where $(t,x)\in \mathbb{R}^+ \times \mathcal{D}$ with $\mathcal{D} \subset \mathbb{R}^d $, $u : = u^h$ is the complex-valued wave function that depends on the time $t$ and the spatial variable $x$, $\Delta$ is the Laplacian operator and $V(x)$ is a given potential. Here $0<h \ll 1$ is the (effective) Planck constant and also called the semiclassical parameter, which serves as the multiscale parameter of the problem. The semiclassical Hamiltonian is $H/h = (A+B)/h = - \frac{h}{2} \Delta + \frac{1}{h} V(x) $. Note that here both the space and time have been rescaled with respect to $h$. The intrinsic time scale is $\tilde{t} = t/h \gg 1$ and the spatial scale is $\tilde{x} = x/h \gg 1$. The intuition is to zoom out to the macroscopic scale in both time and space, while the derivation via non-dimensionalization can be found in, e.g., \cite{LasserLubich2020}.

The semiclassical Schr\"odinger equation arises in a number of applications, in particular, the molecular dynamics under the Born-Oppenheimer approximation~\cite{BornOppenheimer1927}. To be specific, the semiclassical parameter $h$ in this case is the square root of the mass ratio between the electron and nucleus. Equation \eqref{eq:semi_schd} can be derived as the effective description of nuclei under the adiabatic (Born-Oppenheimer) approximation, the mathematical analysis of which can be found in, e.g., \cite{LasserLubich2020,Lubich2008,Hagedorn2007} and papers cited by them. The semiclassical scaling also appears in the quantum-classical molecular dynamics (or the Ehrenfest dynamics), which can be derived mathematically as a partial classical limit of the full molecular Schr\"odinger equation by combining the separation of scales in the wave function and short wave asymptotics~\cite{BornemannNettesheimSchutte1996}. We stress that the constant $h$ may be different from the actual Planck constant, and is simply a small but finite parameter of the model. Mathematically, such systems make up the field of semiclassical analysis.

The main challenges to perform the quantum simulation of this case are of two folds: the high spatial dimensionality, and the oscillations of solution with wavelength $O(h)$ in \textit{both time and space}. 
This suggests that to resolve the solution, the number of spatial grids (or basis) in each dimension is required to be $\Or{(h^{-1})}$ for grid-based methods such as finite difference and Fourier pseudo-spectral spatial discretizations, which is formidable in high dimensions on classical computers. Indeed, this requirement for the spatial grid is made rigorous in, e.g., \cite{BaoJinMarkowich2002,LasserLubich2020}, leading to $\Or(h^{-d})$ total number of spatial grids for $d$-dimensional case.
Therefore, the study of a direct simulation using the Trotter algorithms so far has been mostly restricted to the low-dimensional case, while wavepacket-based methods were preferred when it comes to higher dimensions. However, high spatial dimensionality and hence large Hilbert space are not issues for quantum algorithms. 
Hereafter, we focus on the $h$-scaling of the number of Trotter steps, which is related to the run time. 
The main difficulty for quantum algorithms is the smallness of $h$. More precisely,
the error of existing quantum algorithms~\cite{BerryAhokasCleveEtAl2007,BerryChilds2012,BerryChildsCleveEtAl2014,BerryChildsCleveEtAl2015,BerryChildsKothari2015,LowChuang2017,ChildsMaslovNamEtAl2018,LowWiebe2019,Low2019,ChildsSu2019,ChildsSuTranEtAl2020,AnFangLin2022} is usually measured in the operator norm of the unitary evolution operators, which in turn depends on certain norm of the Hamiltonian. Take one dimension as an example. When taking the number of spatial grids as $\Or{(h^{-1})}$, we denote the spatial discretization of the semiclassical Hamiltonian $H/h = (A+B)/h$
as $A^h + B^h$, where $A^h$ and $B^h$ are the discretized matrices of $- \frac{h}{2} \Delta$ and $\frac{1}{h} V(x)$, respectively, using standard spatial discretization methods such as the finite difference or spectral method which are detailed in \cref{sec:err_dis}. Note that the operator norm spectral norm) of both $A^h$ and $B^h$ are large, namely,
\begin{equation*}
    \norm{A^h} = \Or(h^{-1}), \quad \norm{B^h} = \Or(h^{-1}),
\end{equation*}
and so are the (nested) commutators
\begin{equation*}
    \norm{[A^h,B^h]} = \Or(h^{-1}), \quad \norm{[A^h, [A^h,B^h]]} = \Or(h^{-1}), \quad \norm{[B^h, [A^h,B^h]]} = \Or(h^{-1}).
\end{equation*}
Therefore, post-Trotter algorithms have a query complexity of at least $\Or(h^{-1})$, while the Trotter formulae have commutator scaling~\cite{ChildsSuTranEtAl2020} that results in a smaller power dependence on $h$. When it comes to unbounded operators and separation of scales, the interaction picture framed algorithms~\cite{LowWiebe2019,RajputRoggeroWiebe2021,AnFangLin2022} typically work well in many scenarios, but it is not the case here as
\begin{equation*}
    \norm{H_I(t)} =  \norm{ e^{\I A^h t} B^h e^{-\I A^h t}} = \Or(h^{-1})
\end{equation*}
is on the same scale of $\norm{A^h + B^h}$. 
Recent developments have revealed that the operator norm dependence of the Hamiltonian can be weakened by considering the vector norm or the average sense~\cite{AnFangLin2021,ZhaoZhouShawEtAk2021,ChildsLengEtAl2022,SommaQS2016} for many physical systems. Such techniques typically pose additional assumptions on the initial wave function, but can still result in $h$-dependent query complexity in the semiclassical scenario~\cite{DescombesThalhammer2010,JinLiLiu2021}.
Our work, however, demonstrates that it is possible to eliminate the $h$-dependence in the query complexity for Trotter formulae completely when considering observable error bounds.
That is, when considering the error bounds of the time-evolved observables, the number of time steps required by Trotter formulae can be made $h$-independent. We summarize the comparison of complexity in \cref{tab:trotter_comparison}.

Efficient simulation methods of the semiclassical Schr\"odinger equation have been a long-studied topic in scientific computation and numerical analysis (see \cite{JinMarkowichSparber2011} and \cite{LasserLubich2020} for reviews). \REV{Here we do not aim to provide a review of all the literature. Instead, we will focus on highlighting a few landmark works that are relevant to our study of linear problems and error bounds on observables. The studies of nonlinear problems can be found in, e.g.,\cite{BaoJinMarkowich03,Carles2013,CarlesGallo2017}. I}t was first observed numerically in 2002~\cite{BaoJinMarkowich2002} that certain physical observables can be accurately computed even with a Trotter time step size of $\Or(1)$ -- instead of $\Or(h)$.
Such observable strategy has been successfully applied to various quantum systems over the years (see \cite{LasserLubich2020,JinMarkowichSparber2011} for reviews), together with \REV{asymptotic justification} developed \REV{for quadratic observables via the semiclassical asymptotics. One route is to consider the limiting equation via the Wigner transform, e.g., \cite[Section 5.3]{JinMarkowichSparber2011}. By considering the limit $h \to 0$ one can see that the Trotter formulae preserve its limiting behavior. Unfortunately, the convergence of the Wigner function is in the weak-$\ast$ topology, making the extension of the argument to the spatially discrete setting challenging. More importantly, the precise distance between the actual dynamics and the approximate ones remains an interesting open problem that has attracted} great efforts and progress in the past few years. To estimate the distance for a longer time, another approach via the Husimi function has been proposed.
Denote the Trotter step size as $s$, so that the number of Trotter steps is $t/s$ for the final time $t$. \cite{GolseJin19} investigates the second-order Trotter formulae for the von Neumann equation by measuring the Wasserstein distance between the Husimi functions of the approximate and the exact quantum density operators, and achieves an observable error bound of $\Or(s^2 + h^{1/2})$ for the second-order splitting, and a uniform-in-$h$ bound is also obtained in \cite{GolseJin19} as $\Or(s^{2/3})$ under suitable assumptions of the initial wave function that reduces the accuracy of the second-order algorithm to an order of $2/3$. The first bound for the semiclassical Schr\"odinger equation is given in the breakthrough work \cite{LasserLubich2020} by exploring the properties of Husimi functions and the St\"ormer-Verlet integrator. For the second-order Trotter formula, it provides a tighter bound of $\Or(s^2 + h^{2})$ for observables that are quantizations of Schwartz functions (that is, smooth functions whose derivatives, including the function itself, decay at infinity faster than any power) in $\mathbb{R}^n$. The extension of such observable expectation bounds to quantum-classical molecular dynamic, a weakly nonlinear case, has been addressed in~\cite{FangTres2021}, improving the uniform bound estimate to $\Or(s^{4/3})$ under suitable assumptions of the initial wave function. \REV{On a different note, in certain nonlinear systems, the WKB-based analysis can yield a uniform estimate for the position and current densities when the initial condition is in a WKB form~\cite{Carles2013,CarlesGallo2017}. Nevertheless, it is important to acknowledge that, similar to other WKB-based analysis, this type of estimation is not globally valid over time.}
All observable analysis focuses on the spatially continuous case.

With the additive error bound, however, the algorithm only works when a low precision is needed, namely, the precision $\epsilon > \Or(t h^2)$ as is pointed out in \cite{JinLiLiu2021}. In other words, this means that the (additive) error bound (e.g., $\Or(s^2 + h^2)$) of the Trotter formulae does \textit{not} vanish even as the time step size $s$ shrinks to zero (i.e. the number of Trotter steps goes to infinity), while the actual error of Trotter formulae should diminish. This seemingly subtle issue is non-negligible, as it implies the performance of the Trotter formulae can not be improved beyond certain precision even taking the number of Trotter steps (and hence computational cost) to be infinity. Note that the $h$ part in the additive error bound is an asymptotic error coming from estimating the macroscopic limit of the Trotter formulae in the proving strategy. However, the Trotter formulae implemented are for the Schr\"odinger equation (microscopic description) without any prior asymptotic approximation, and hence should be free of asymptotic errors.
Also for a given problem in practice,  $h$ is a small but finite parameter. To account for high precision as well as closing the gap of the asymptotic error, a uniform (i.e. independent of $h$) error bound is desired.

\vspace{1em}
\noindent \textbf{Main Questions:}

Despite remarkable progress, there are still two interesting and challenging aspects remaining in this problem: to prove a uniform-in-$h$ observable error bound \REV{for the linear Schr\"odinger equation} without compromising the convergence order of the Trotter formulae so that high precision can be achieved, and to perform the observable analysis in the spatially discretized setting.
Our paper aims to address both aspects. Note that besides the uniformity, the second aspect is also of importance here, as algorithms are implemented with both time and spatial discretizations and hence it is important to understand the behavior in the fully discretized setting.
 
In terms of the tools of analysis, we adopt the techniques of semiclassical and microlocal analysis, which are well-established mathematical subjects in the study of partial differential equations, particularly in quantum dynamics (see textbooks~\cite{zworski_book,RobertCombescure2021,Martinez2002}). Its application to quantum algorithms was first introduced in \cite{AnFangLin2022} achieving a surprising superconvergence result for the quantum highly oscillatory protocol (qHOP) by a careful investigation of the pseudo-differential operators. Here we use both standard semiclassical Weyl calculus and discrete microlocal analysis~\cite{Borns-Weil2022,ChristiansenZworski2010,DyatlovJezequel2021,FaouGrebert2021}, the latter of which provides a nice correspondence between discrete matrices and the quantization of phase-space functions on the torus. This allows us to extend our continuous-in-space observable results (in \cref{sec:err_cts}) to the case with spatial discretization (in \cref{sec:err_dis}). We provide brief yet self-contained introductions to both topics in \cref{ss:microlocal} and \cref{sec:discrete_microlocal}.

\begin{table}[]
    \centering
    \begin{tabular}{p{3.5cm}|p{5.2cm}|p{4cm}}\hline\hline
        Error Measurement & Work/Method & Query complexity in the Planck constant $h$  \\\hline
       \multirow{4}{8em}{Unitaries (operator norm)} & 
       $p$-th order Trotter~\cite[Theorem 6]{ChildsSuTranEtAl2020} & $\Or(h^{-1/p})$  \\
       \cline{2-3} & Truncated Taylor series~\cite{BerryChildsCleveEtAl2014,KivlichanWiebeBabbushEtAl2017} & $\widetilde{\Or}(h^{-1})$  \\
       \cline{2-3} & QSVT~\cite{GilyenSuLowEtAl2018} & ${\Or}(h^{-1})$  \\
       \cline{2-3} & Interaction picture~\cite{LowWiebe2019} & $\widetilde{\Or}(h^{-1})$  \\\hline
       \multirow{2}{10em}{Wave functions \newline (for initial data with good regularity)} & $p$-th order Trotter ~\cite{DescombesThalhammer2010,JinLiLiu2021,ChildsLengEtAl2022} & $\Or(h^{- 1/p})$  \\ 
       \cline{2-3} &  rescaled Dyson series in the interaction picture~\cite{ChildsLengEtAl2022} & $\widetilde{\Or}(h^{-1})$   \\
       \hline
       \multirow{1}{10em}{Observable evolution} 
       & This work (Trotter1\&2) & ${\Or}(1)$ \\\hline\hline
    \end{tabular}
    \caption{Comparison of query complexity estimates for simulating the semiclassical Schr\"odinger equation~\eqref{eq:semi_schd} using second order Trotter method or higher order Trotter or post-Trotter quantum algorithms.
    The query complexity is measured by the number of required Trotter steps for Trotter-type methods, or the query complexity under the standard query model for post-Trotter methods. The simulation time $t$ is $\Or(1)$. `This work' refers to the error of either the time-evolved observable in the operator norm or the observable expectation using either the first-order or the second-order Trotter formula for the general observables in the symbol class. Throughout the paper $f=\wt{\Or}(g)$ if $f=\Or(g\operatorname{polylog}(g))$. See \cite{thesupplement} for details of the derivation of the scalings. }
    \label{tab:trotter_comparison}
\end{table}

\vspace{1em}
\noindent\textbf{Contribution:}

In this paper, we establish the first rigorous proof of the \textit{uniform-in-$h$} observable error bounds for both the first-order and second-order Trotter formulae \textit{without} sacrificing their orders of accuracy. This 
gives rise to an $h$-independent query complexity for Trotter formulae that vastly improves the state-of-the-art error bounds in either the operator norm or the vector norm. It is worth noting that the usual estimating strategy involving a multiscale parameter typically studies the macroscopic limit of the numerical schemes (in this case, the symplectic integrators as the limit of the Trotter formulae) and uses it as a stepping stone. To be specific, the total error is typically estimated as a sum of three contributions -- the error between the actual algorithm and its macroscopic limit, the error between the actual differential equation (Sch\"odinger equation) and its macroscopic limit, and the error between the microscopic limit of the differential equation and the microscopic limit of the numerical algorithm. This will inevitably make the error bounds additive, instead of uniform.  
Here we manage to conduct the error estimate directly at the microscopic level, that is, focusing on the error between the Trotter formulae and the Schr\"odinger equation themselves. Our analysis works with the errors in the operator norm of time-evolved observables, that is,
\begin{equation*}
   T(t) := e^{\I H t/h} O e^{-\I H t/h}
\end{equation*}
for an observable $O$. Hence, it measures the worst-case scenario for any given initial wave function and does not impose extra assumptions on the initial wave function. The observable expectation for a given initial wave function can be achieved as an immediate corollary. It is also worth mentioning that our results are shown for any (semiclassical) Weyl quantized operators $A$ and $B$ (as defined in \eqref{eq:weylquant}) that need not to be $-\frac{h^2}{2}\Delta$ and $V(x)$, which are quantizations of the kinetic energy $p^2/2$ and the potential $V(x)$, as in the case of the semiclassical Schr\"odinger equation. Note that the rule of Weyl quantization is consistent with the quantization used in physics, in the sense that the quantization of the momentum $p$ is $-\I h \nabla$ and the quantization of the position $x$ is a multiplication operator by $x$. Our result also applies to other quantized operators, such as polynomials of the momentum operators and $V(x)$. Besides, it is also interesting to observe that our result generalizes the type of observables as a byproduct. 
The analysis works with a larger symbol class $S$ (to be defined in \cref{ss:microlocal}) that does not require the decay of derivatives compared to the Schwartz class. As a simple example, the position observable $\cos(x)$ does not belong to the Schwartz class, but belongs to the symbol class.

As a second contribution, we derive the observable error bound in the spatially discrete setting by taking advantage of the discrete microlocal analysis on the quantized torus that corresponds to finite dimensional matrices. This is, to our knowledge, the first application of the discrete microlocal analysis in quantum computation.
This allows one to give a rigorous complexity analysis of the Trotter formulae for observable calculations, as the quantum algorithms implement the matrices induced by the spatial discretization, instead of the continuous unbounded operators. In literature, the relaxed (larger) Trotter time step size has been only observed numerically and recognized when the Fourier pseudo-spectral spatial discretization is applied. Interestingly, we show that an $h$-independent time step size is also permitted for the finite difference spatial discretization.

\vspace{1em}
\noindent\textbf{Related work:}

The Hamiltonian simulation problem involving the unbounded operator $\Delta$ has been referred to as the real-space Hamiltonian simulation, Hamiltonian simulation with unbounded operators or in the first quantization~\cite{AnFangLin2021,ChildsLengEtAl2022,SuBerryWiebeEtAl2021,KivlichanWiebeBabbushEtAl2017,TongAlbertEtAl2021,BabbushBerryEtAl2019}.
The error analysis of the Trotter formulae has been a widely studied topic. Childs et al. \cite{ChildsSuTranEtAl2020} provide the commutator scaling for high-order Trotter formulae that significantly improves the state-of-the-art error bounds for bounded operators yielding near-best asymptotic complexities for simulating problems such as $k$-local Hamiltonians. More recent improvement for the bounded case of Trotter-type formulae and Hamiltonian simulation algorithms can be found in, e.g., \cite{SahinogluSomma2021,ZhaoZhouShawEtAk2021,McArdleCampbellSu2022,SuHuangCampbell2021,TranSuEtAl2021,YiCrosson2022,LowSuTongTran2023,ChenBrandao2024,zhao2024entanglement,yu2024observabledriven}. When unbounded operator is involved, An–Fang–Lin \cite{AnFangLin2021} establish the vector norm error bound for both the standard and the generalized Trotter formulae of the first and second orders. By taking into account the initial wave function, the vector norm analysis considers the error in the wave function, instead of the unitary evolution operator, which weakens the operator norm dependence of the Hamiltonian and vastly reduces the overhead caused by the spatial discretization. The wave function errors for high-order Trotter formulae and post-Trotter methods have been carefully studied in \cite{ChildsLengEtAl2022}.

For bounded Hamiltonians with geometric local structures, it has been found that improved performance and error estimates can be achieved for the simulation of local observables. Taking advantage of the locality, Lieb–Robinson bounds can be estabilished for Hamiltonians such as power-law interactions~\cite{HaahHastingsEtAl2018,TranGuoEtAl2019,TranEhrenbergEtAl2019,TranChenEtAl2020,TranGuoEtAl2021,TranGuoDeshpandeEtAl2021}. In contrast, our work focuses on the Schr\"odinger equation in the real space, whose Hamiltonian is not geometrically local. Our observable error bound is achieved by the virtue of the mechanism from semiclassical analysis.

\vspace{1em}
\noindent\textbf{Organization:}

The rest of this paper is now organized as follows: In \cref{sec:prelim} we give a brief yet self-contained overview of semiclassical Weyl calculus and some elementary lemma that play an important role in later analysis. 
\cref{sec:err_cts} focuses on Trotter errors for the observables for both short time and long time, while keeping the spatial degrees of freedom continuous. We provide uniform-in-$h$ error bounds in the operator norm for the time-evolved observables, which immediately imply uniform-in-$h$ error bounds for the observable expectations. These results are used to establish the observable error bounds in the spatially discrete case. \cref{sec:err_dis} discusses two types of spatial discretizations and their connection to the quantized torus, and then provides the observable error analysis for the Trotter formulae in the spatial discretized setting, using discrete microlocal analysis. Numerical results are provided in \cref{sec:numerics} to verify our estimates. Finally, in \cref{sec:conclusion}, we conclude with some further remarks.

\section{Preliminary} \label{sec:prelim}

In this section, we discuss a number of preliminaries as the main tools that we use in our estimates. In particular, we give a short yet self-contained review of traditional semiclassical symbol calculus, which plays a crucial role in proving our results. Variation of parameter lemma is also discussed.

\subsection{Semiclassical Weyl calculus}\label{ss:microlocal}

We review semiclassical microlocal analysis on $\RR^d$, which will be our main tool in deriving observable-norm bounds. Such analysis is concerned with $h$-dependent operators, where $h\in(0,1)$ is a \emph{semiclassical parameter} that we take to be small. For a more extensive introduction see the books by Zworski~\cite{zworski_book} \REVV{and by Martinez~\cite{Martinez2002}}.

To introduce the notion of the symbol class, we begin by revisiting the definition of an order function. A measurable function $m: \mathbb{R}^{2d} \to (0, \infty) $ is called an order function if there exist constants $C$, $N$ such that $m(w)\leq C \langle z-w\rangle^N m(z)$ for all $w,z \in \mathbb{R}^{2d}$.
Some common examples are
\begin{equation}
    m(z) \equiv 1, \quad m(z) = \langle \xi \rangle = (1+|\xi|^2)^{1/2}, \quad m(z) = \langle x \rangle^\alpha \langle \xi \rangle^\beta ,
\end{equation}
for $z = (x, \xi)$ and $\alpha, \beta \in \mathbb{R}$. 
The space of \emph{symbols}, also known as the symbol class, is denoted by $S(m)$ and defined as follows: 
$$S(m)=S(m;\RR^{2d}):=\left\{a\in\cinf(T^*\RR^d):\left|\dd_x^{\alpha}\dd_{\xi}^{\beta}a(x,\xi)\right|\le C_{\alpha\beta} m \right\}.$$ 
Here $T^*$ stands for the cotangent bundle. We remark that such symbols are allowed to depend on a semiclassical parameter $h$, as long as the constants $C_{\alpha\beta}$ remain $h$-independent. The definition of a symbol is designed to be quantized into an operator, which we now demonstrate. \REVV{Given $a\in S(m)$}, define its \emph{Weyl quantization} as $\op(a)=\op_h(a): \Sc(\RR^d)\to\Sc(\RR^d)$ given by \begin{equation}\label{eq:weylquant}
(\op(a)u)(x):=\frac{1}{(2\pi h)^d}\iint_{\RR^{2d}} e^{\frac{\I}{h}\la x-y,\xi\ra}a\left(\frac{x+y}{2},\xi\right)u(y)\,dy\,d\xi,
\end{equation} 
where ${\displaystyle {\mathcal {S}}(\mathbb {R} ^{d})}$ denotes the Schwartz space defined as the space of all smooth functions acting on $\mathbb {R} ^{d}$ that are rapidly decreasing at infinity along with all partial derivatives. By distributional pairing, we may extend $\op_h(a)$ to act on the class of tempered distributions $\Sc'(\RR^d)$. 
Such operators are called semiclassical pseudodifferential operators, and are self-adjoint on $L^2$ provided that $a$ is real-valued. Weyl quantization can be defined for more general symbol classes such as $S_\delta(m)$ and $S_\delta^{m,k}(T^*\mathbb{R}^d)$ (see, e.g., \cite{zworski_book,NonnenmacherSjostrandZworski2014}). For simplicity, we work with $S(m)$. We also denote $S(1)$ as $S$, which will be the symbol class for the observables. We remark that if $a(x,\xi)=a(x)$, then $\op_h(a)u(x)=a(x)u(x)$, and if $a(x,\xi)=a(\xi)$, then $\mathcal{F}_h\op_h(a)u(\xi)=a(\xi)\mathcal{F}_hu(\xi)$, where $\mathcal{F}_h$ is the \emph{semiclassical Fourier transform}
\begin{equation}\label{eq:scft}\mathcal{F}_hu(\xi):=\frac{1}{(2\pi h)^{\frac{d}{2}}}\int_{\RR^d} e^{-\frac{\I}{h}\la x,\xi\ra}u(x)\,dx.\end{equation} 
Note that this agrees with the quantization rules in physics, where $\op_h(x) = \hat x$ is the multiplication operator by $x$ and $\op_h(p) = \hat p = -\I h \nabla_x$. 
In these special cases, the observables are simply functions of position and of momentum, respectively. The general formula (\ref{eq:weylquant}) serves as a recipe for quantization of functions involving both position and momentum together.

One of the first results from pseudodifferential calculus is the \emph{composition rule}, which states that the composition $\op_h(a)\op_h(b)$ is itself a semiclassical pseudodifferential operator $\op_h(a\#b)$, where 
\begin{equation}\label{eq:comp}a\#b=ab+\frac{h}{2\I}\{a,b\}+h^2r.\end{equation} Here $\{a,b\}$ is the Poisson bracket $\{f,g\}:=\la\dd_{\xi}f,\dd_xg\ra-\la\dd_xf,\dd_{\xi}g\ra$.
\REVV{For $a \in S(m_1)$ and $b \in S(m_2)$, $a \# b \in S(m_1 m_2)$.
} 
Intuitively, the first term of (\ref{eq:comp}) says that composing operators is equivalent to multiplying their symbols ``up to order $h$."

From (\ref{eq:comp}), we may also derive a formula for the commutator of two Weyl quantizations. It may be shown that an extra order of $h^2$ cancels in the remainder to give 
\begin{equation}\label{eq:comm}
[\op_h (a),\op_h (b)]=\op_h\left(\frac{h}{\I}\{a,b\}+h^3r\right)
\end{equation} 
Another helpful fact is that when either $a$ or $b$ is a quadratic function, the remainder $r$ becomes $0$.

A celebrated result of Calderon and Vaillancourt (see \cite[Chapters 4 and 13]{zworski_book}) gives that for $a\in S(1)$, $\op_h(a):L^2(\RR^d)\to L^2(\RR^d)$ is bounded uniformly in $h$. In fact, one has 
\begin{equation}\label{eq:cv}
\|\op_h(a)\|_{L^2(\RR^d)\to L^2(\RR^d)}\le\|a\|_{L^{\infty}}+C(a)h
\end{equation}
where $C(a)$ depends on finitely many derivatives of $a$ but is independent of $h$. 

The final result we will need is a theorem of Egorov, which (in the form we will use) gives for symbols $a, b$ and $t\le T$, that
\begin{equation}\label{eq:egorov}
    e^{\frac{\I t}{h}\op_h(b)}\op_h(a)e^{-\frac{\I t}{h}\op_h(b)}=\op_h \left(\phi_t^*a\right)+h^2\op_h(r)
\end{equation} where $\phi_t$ is the flow generated by the classical Hamiltonian vector field $H_b=\la\dd_{\xi}b,\dd_x\ra-\la\dd_xb,\dd_{\xi}\ra$. \REVV{When $a, b \in S(1)$, one has $r\in S(1)$ with $\|\op_h( r)\|\le Ct$ for some $C$ that depends on $T$ and on the derivatives of $a$ and $b$ but is independent of $t$ and $h$. Another important case of the Egorov theorem is when $b$ is the Hamiltonian for harmonic oscillators, that is $b = p^2 + V(x)$ with a quadratic $V$. In this case, $r$ becomes zero, making the Egorov theorem exact.} This theorem can also be constructed for general symbols. For a proof see \REVV{\cite[Chapter 4]{Martinez2002}}, \cite[Section 6]{LasserLubich2020} or \cite[Chapter 11]{zworski_book}.

Egorov's theorem makes precise the ``classical-quantum correspondence," by showing that (for a fixed time at least) the quantum evolution of an observable in the Heisenberg picture follows the corresponding classical Hamiltonian evolution up to order $h$.

\subsection{Elementary lemma}
We review the variation of parameter lemma for the Sylvester-type equations \cite[Lemma1]{AuzingerKochThalhammer2014} which will be used in later proofs. 
\begin{lemma}[Variation of parameter formula] \label{lem:var_para}
Let $A$ and $B$ be time-independent operators and $G$ a time-dependent inhomogeneity. Consider the inhomogeneous Sylvester equation of $X(t)$ given by
\begin{equation} \label{eq:sylvester}
\frac{d}{dt} X = X A + BX + G,
\end{equation}
with initial condition given as $X(0)$. The solution admits the representation
\begin{equation} \label{eq:var_para}
    X(t) = e^{tB} X(0) e^{tA} + \int_0^t e^{(t-s) B} G(s) e^{(t-s) A}\, ds.
\end{equation}
\end{lemma}

\section{Trotter error analysis for time-evolved observables} \label{sec:err_cts}
In this section, we prove observable error bounds when measuring in the operator norm sense (worst-case scenario) for both short-time evolution and the long-time ones. We also show that using the operator norm observable error result, one immediately gets the error bound in terms of the observable expectations. These error bounds give the estimate of the number of Trotter steps needed.

Consider the first-order and second-order Trotter formulae and a Hamiltonian in the semiclassical regime $H^h := H/h = (A + B)/h$. Denote the time step size as $s$. The first-order Trotter formula (i.e.\ Lie-Trotter splitting) is
\begin{equation} \label{eq:trotter1}
    U_1(t+s,t) = e^{ -\I B s/h} e^{ -\I A s/h},
\end{equation}
and the second-order Trotter formula (i.e.\ Strang splitting) is
\begin{equation}\label{eq:trotter2}
    U_2(t+s,t) = e^{-\I B s/(2h)} e^{-\I As/h}e^{-\I B s/(2h)}.
\end{equation}

\subsection{Short-time evolution error}
We first consider the short-time error of the observables for the first-order Trotter formula. The idea is to directly investigate the evolution operator of the observable. In this section, we fix Hamiltonian $H=A+B$ where $A=\op_h(a), B=\op_h(b) = \op_h(V)$. \REVV{For simplicity, we assume $V(x) \in S = S(1)$ in the case of the semiclassical Schr\"odinger equation, and we also consider the observable is the symbol class $S$. We remark that the proof of \cref{thm:trot1} and \cref{thm:trot2} may be extended for more general symbol classes. Here we focus on $S$ for simplicity. We also recall that $a = \xi^2$ is a quadratic function so that the corresponding Egorov theorem is exact as discussed in the preliminary.}

\REVV{Another motivating example of $a$ is $2-2\cos(2\pi\xi)$. This is because as will be discussed in \cref{sec:discrete_microlocal}, the Laplacian matrix, after spatial discretization using the finite differences, can be identified with the Weyl quantization of the function $2-2\cos(2\pi\xi)$ on the quantized torus. \cref{lem:H_N_L^2_bound} further shows that the norm of such discrete quantization can be bounded by the $L^2 \to L^2$ norm of its continuous counterpart. 
Although the Egorov theorem is no longer exact in this case, this function belongs to the symbol class $S(1)$ so that the following proofs still hold.}

The aim of this session is to set up the stage so that we can apply such theorems to discrete operators in \cref{sec:discrete_microlocal}. \cref{thm:trot1} holds for both the case of $a = \xi^2$ and $a \in S$, while our current proof for \cref{thm:trot2} only considers $a \in S$. This is motivated by the discretization process and will allow us to apply the result to the fully discretized case in \cref{sec:discrete_microlocal}.

 \begin{theorem}[Local error of observable for Trotter1]\label{thm:trot1} Let $O = \op_h(o)$ be the Weyl quantization of a function $o(x,p) \in S$.
 Denote its time evolution under the exact dynamics as
\begin{equation}\label{eq:ob_exact}
    T(s): = e^{\I Hs/h} O e^{-\I Hs/h},
\end{equation}
and under the first-order Trotter formula as
\begin{equation}\label{eq:ob_exact}
    T_1(s): = e^{\I As/h} e^{\I Bs/h} O e^{-\I Bs/h} e^{-\I As/h}. 
\end{equation}
The difference between the two can be estimated as
\begin{equation}\label{eq:ob_err_trotter1}
    \norm{ T_1(s) - T(s)}_{L^2 \to L^2} \leq C \left( s^2 + s^2 h \right).
\end{equation}
\end{theorem}
\begin{proof}
Taking the derivative of $T_1(s)$ with respect to $s$ yields
\begin{equation} \label{eq:ds_T1}
    \partial_s T_1 = -\frac{\I}{h} T_1 A + \frac{\I}{h} A T_1 + \frac{\I}{h} e^{\I As/h} e^{\I Bs/h} [B,O] e^{-\I Bs/h} e^{-\I As/h}.
\end{equation}
Similarly, the derivative of $T(s)$ reads
\begin{equation} \label{eq:ds:T}
    \partial_s T = \frac{\I}{h} [H, T].  
\end{equation}
Subtracting the two equations, we find that the difference $X: = T_1 - T$ satisfies an inhomogeneous generalized Sylvester equation
\begin{equation*}
    \partial_s X = \frac{\I}{h} H X - \frac{\I}{h} X H +  g(s),
\end{equation*}
where \begin{equation*}
    g(s) = \frac{\I}{h} [T_1, B] - \frac{\I}{h} e^{\I As/h} e^{\I Bs/h} [O,B] e^{-\I Bs/h} e^{-\I As/h}.
\end{equation*}
Thanks to the variation of parameter formula to the Sylvester-type equation (\cref{lem:var_para}), we have
\begin{equation*}
    X(s) = e^{\I H s/ h} X(0) e^{-\I H s/h} + \int_0^s e^{\I H (s - \tau)/ h} g(\tau) e^{-\I H (s-\tau) /h}d\tau,
\end{equation*}
and hence it suffices to estimate $\norm{g(\tau)}_{L^2 \to L^2}$ since $X(0) = 0$. We rewrite $g$ as
\begin{align*}
g(s)&=\frac{\I}{h}e^{\frac{\I As}{h}}e^{\frac{\I Bs}{h}}\left(Oe^{\frac{-\I Bs}{h}}e^{\frac{-\I As}{h}}Be^{\frac{\I As}{h}}e^{\frac{\I Bs}{h}}-e^{\frac{-\I Bs}{h}}e^{\frac{-\I As}{h}}Be^{\frac{\I As}{h}}e^{\frac{\I Bs}{h}}O-OB+BO\right)e^{\frac{-\I Bs}{h}}e^{\frac{-\I As}{h}}\\
&=\frac{\I}{h}e^{\frac{\I As}{h}}e^{\frac{\I Bs}{h}}[O,E]e^{\frac{-\I Bs}{h}}e^{\frac{-\I As}{h}},
\end{align*}
where 
\[
E=e^{\frac{-\I Bs}{h}}e^{\frac{-\I As}{h}}Be^{\frac{\I As}{h}}e^{\frac{\I Bs}{h}}-B.
\]
But by Egorov's theorem (\ref{eq:egorov}) we have $E=\op_h(\phi_s^*b-b)+R$ where $\|R\|_{L^2\to L^2}=\Or(sh^2)$, and $\phi_s^*b-b$ has all of its derivatives bounded by $\Or(s)$.
Then the commutator formula (\ref{eq:comm}) together with the Calderon-Vaillancourt theorem (\ref{eq:cv}) gives
\[
\|g(s)\|=\left\|\frac{1}{h}[O,E]\right\|=\Or_{L^2\to L^2}(s+sh).
\]
Therefore, the local truncation error is
\begin{equation*}
    \norm{X(s)} \leq s \sup_{\tau \in [0, s]} \norm{g(\tau)} \leq C\left( s^2 + s^2h \right),
\end{equation*} as desired.
\end{proof}

The short-time error of the observables for the second-order Trotter formula can be summarized as follows.
\begin{theorem}[Local error of observable for Trotter2]\label{thm:trot2} Let $O = \op_h(o)$ be the Weyl quantization of $o\in S$. Denote its time evolution under the exact dynamics as
\begin{equation}\label{eq:ob_exact}
    T(s): = e^{\I Hs/h} O e^{-\I Hs/h},
\end{equation}
and under the second-order Trotter formula as
\begin{equation}\label{eq:ob_exact}
    T_2(s): = e^{\I As/(2h)} e^{\I Bs/h}e^{\I As/(2h)} O e^{-\I As/(2h)}e^{-\I Bs/h} e^{-\I As/(2h)}. 
\end{equation}
The difference between the two can be estimated as
\begin{equation}\label{eq:ob_err_trotter2}
    \norm{ T_2(s) - T(s)}_{L^2 \to L^2} \leq C( s^3 + s^3 h).
\end{equation}
\end{theorem}
\begin{proof}
The proof proceeds as in Theorem \ref{thm:trot1}. We have 
$$\begin{gathered}\dd_sT_2(s)=\frac{\I}{h}\left(\frac{A}{2}T_2+e^{\frac{\I As}{2h}} Be^{\frac{\I Bs}{h}}e^{\frac{\I As}{2h}} O e^{-\frac{\I As}{2h}}e^{-\frac{\I Bs}{h}} e^{-\frac{\I As}{2h}}+e^{\frac{\I As}{2h}} e^{\frac{\I Bs}{h}}\frac{A}{2}e^{\frac{\I As}{2h}} O e^{-\frac{\I As}{2h}}e^{-\frac{\I Bs}{h}} e^{-\frac{\I As}{2h}}\right.\\
\left.\quad-e^{\frac{\I As}{2h}} e^{\frac{\I Bs}{h}}e^{\frac{\I As}{2h}} O e^{-\frac{\I As}{2h}}\frac{A}{2}e^{-\frac{\I Bs}{h}} e^{-\frac{\I As}{2h}}-e^{\frac{\I As}{2h}} e^{\frac{\I Bs}{h}}e^{\frac{\I As}{2h}} O e^{-\frac{\I As}{2h}}e^{-\frac{\I Bs}{h}}B e^{-\frac{\I As}{2h}}-T_2\frac{A}{2}\right).
\end{gathered}$$

Letting $X(s)=T(s)-T_2(s)$, we compute that
$$\dd_sX=\frac{\I}{h}[A+B,X]+\frac{\I}{h}\left[\wt G,T_2\right]$$
where $$\wt G(s)=\frac{A}{2}+B-e^{\frac{\I As}{2h}}Be^{-\frac{\I As}{2h}}-e^{\frac{\I As}{2h}}e^{\frac{\I Bs}{h}}\frac{A}{2}e^{-\frac{\I Bs}{h}}e^{-\frac{\I As}{2h}}.$$
By variation of parameter formula (\cref{lem:var_para}) \begin{equation}\label{eq:var_par2}X(s)=\int_0^se^{\frac{\I}{h}(A+B)(s-\tau)}\frac{\I}{h}\left[\wt G(\tau),T_2(\tau)\right]e^{-\frac{\I}{h}(A+B)(s-\tau)}d\tau.
\end{equation} We seek to bound $\wt G$. We let $$G(s)=\frac{1}{2}\left(A-e^{\frac{\I}{h}Bs}Ae^{-\frac{\I}{h}Bs}\right)+e^{-\frac{\I As}{2h}}Be^{\frac{\I As}{2h}}-B$$ so $\wt{G}=e^{\frac{\I As}{2h}}Ge^{-\frac{\I As}{2h}}$. The fundamental theorem of calculus gives $$G(s)=\int_0^se^{\frac{\I A\tau}{2h}}Qe^{-\frac{\I A\tau}{2h}}-e^{-\frac{\I B\tau}{h}}Qe^{\frac{\I B\tau}{h}}d\tau$$ where $Q=\frac{\I}{2h}[A,B]$ is a semiclassical pseudodifferential operator quantizing $q\in S$.

By Egorov's theorem (\ref{eq:egorov}),
$G(s)=\int_0^s\op_h(\psi^*_{\tau}(q)-\phi^*_{\tau}(q))+R(\tau)\,d\tau$ where $\|R(\tau)\|_{L^2\to L^2}=\Or(\tau h^2)$ which gives $$G(s)=\op_h\left(\int_0^s\psi^*_{\tau}(q)-\phi^*_{\tau}(q)\,d\tau\right)+R_1(s)$$ with $\|R_1(s)\|_{L^2\to L^2}=\Or(s^2h^2)$. Then
\begin{align*}
    \frac{\I}{h}\left[\wt G,T_2\right]&=\frac{\I}{h}e^{-\frac{\I As}{2h}}Ge^{\frac{\I As}{2h}}T_2-T_2e^{-\frac{\I As}{2h}}Ge^{\frac{\I As}{2h}}\\
    &=\frac{\I}{h}e^{-\frac{\I As}{2h}}\left(\left[G,e^{-\frac{\I As}{2h}}T_2e^{-\frac{\I As}{2h}}\right]\right)e^{-\frac{\I As}{2h}}.
\end{align*}

But the symbol $\int_0^s\psi^*_{\tau}(q)-\phi^*_{\tau}(q)\,d\tau$ of $G$ has all its derivatives bounded by $\Or(s^2)$, so once again by the commutator formula (\ref{eq:comm}) and Calderon-Vaillancourt theorem (\ref{eq:cv}) we have
\begin{equation}\label{eq:comm_bound2}\left\|\frac{\I}{h}\left[\wt G,T_2\right]\right\|=\Or_{L^2\to L^2}(s^3+hs^3).\end{equation}
Then (\ref{eq:ob_err_trotter2}) follows from (\ref{eq:comm_bound2}) and (\ref{eq:var_par2}).
\end{proof}

Though we keep the $h$ contributions in both \cref{thm:trot1} and \cref{thm:trot2}, we see that unlike the operator norm error or the vector norm error, the $h$ is no longer in the denominator. Therefore in the physically relevant regime $0< h \leq 1$, we immediately obtain a uniform-in-$h$ error bound.
\begin{corollary} \label{cor:error_short_time}
Let $O = \op_h(o)$ be the Weyl quantization of $o\in S$ and $0<h \leq 1$. The observable errors for the short-time evolution of the first-order and the second-order Trotter formulae are bounded by
\begin{align}
    \label{eq:unif_c_b1}\norm{ T_1(s) - T(s)}_{L^2 \to L^2} &\leq C s^2,
    \\
    \label{eq:unif_c_b2}\norm{ T_2(s) - T(s)}_{L^2 \to L^2} &\leq C s^3,
\end{align}
for some constant $C$ independent of $s$ and $h$.
\end{corollary}

\subsection{Long-time evolution error and observable expectation}

We can now give bounds on the long-time Trotter error in the observable norm. We state the result for first and second-order Trotter formulae, but prove it only for second-order as the proof for first-order is identical using (\ref{eq:unif_c_b1}) instead of (\ref{eq:unif_c_b2}) in Corollary \ref{cor:error_short_time}.

\begin{theorem}\label{thm:long_trot1}
Let the observable $O = \op_h(o)$ be the Weyl quantization of a symbol $o\in S(\RR^{2d})$. Let $t=ns$ and $0<h\le 1$, and let $$T(t):=e^{\I Hs/h}Oe^{-\I Hs/h}.$$ and $$T_{1,n}(s):=(e^{\I As/h} e^{\I Bs/h})^n O (e^{-\I As/h}e^{-\I Bs/h})^n.$$ Then the global error can be estimated as
$$\|T_{1,n}(s)-T(t)\|_{L^2\to L^2}\le C_t s$$
for some $C_t$ independent of $n$ and $h$.
\end{theorem}

\begin{theorem}\label{thm:long_trot2}
Let the observable $O = \op_h(o)$ be the Weyl quantization of a symbol $o\in S(\RR^{2d})$. Let $t=ns$ and $0<h\leq 1$, and let $$T(t):=e^{\I Hs/h}Oe^{-\I Hs/h}.$$ and $$T_{2,n}(s):=(e^{\I As/(2h)} e^{\I Bs/h}e^{\I As/(2h)})^n O (e^{-\I As/(2h)}e^{-\I Bs/h} e^{-\I As/(2h)})^n.$$ Then the global error can be estimated as
$$\|T_{2,n}(s)-T(t)\|_{L^2\to L^2}\le C_t s^2$$
for some $C_t$ independent of $n$ and $h$.
\end{theorem}
\begin{proof}[Proof of Theorem \ref{thm:long_trot2}]
We have 
\begin{align*} & \|T_{2,n}(s)-T(t)\|_{L^2\to L^2} 
\\
\le & \|T_{2,n}(s)-e^{\I As/(2h)} e^{\I Bs/h}e^{\I As/(2h)} T((n-1)s)e^{-\I As/(2h)}e^{-\I Bs/h} e^{-\I As/(2h)}\|_{L^2\to L^2}\\&\quad+\|e^{\I As/(2h)} e^{\I Bs/h}e^{\I As/(2h)} T((n-1)s)e^{-\I As/(2h)}e^{-\I Bs/h} e^{-\I As/(2h)}-T(t)\|_{L^2\to L^2}\\
\le & \|T_{2,n-1}(s)-T((n-1)s)\|_{L^2\to L^2}\\&\quad+\|e^{\I As/(2h)} e^{\I Bs/h}e^{\I As/(2h)} T((n-1)s)e^{-\I As/(2h)}e^{-\I Bs/h} e^{-\I As/(2h)}-T(t)\|_{L^2\to L^2}.
\end{align*}
Then by induction we get 
\begin{multline*}
    \|T_{2,n}(s)-T(t)\|_{L^2\to L^2} \\ \le\sum_{j=0}^{n-1}\|e^{\I As/(2h)} e^{\I Bs/h}e^{\I As/(2h)} T(js)e^{-\I As/(2h)}e^{-\I Bs/h} e^{-\I As/(2h)}-T((j+1)s)\|.
\end{multline*}
But by Egorov's theorem, $T(js)$ is a pseudodifferential operator with symbol in $S$ whose seminorms are bounded by a constant depending on $t$ (uniform in $j$). Then applying (\ref{eq:unif_c_b2}) in Corollary \ref{cor:error_short_time} to each term and summing gives $$\|T_{2,n}(s)-T(t)\|_{L^2\to L^2}\le C_t s^2.$$
as was to be shown.
\end{proof}

\begin{remark}
The error bounds in \cref{thm:long_trot1} and \cref{thm:long_trot2} are uniform in $h$, where $0 <h \leq 1$. When $h > 1$, though less physically relevant here, one can directly use the operator norm error in terms of the evolution operators or the vector norm error in the wave function. 
\end{remark}

It is worth pointing out that the operator norm error bound immediately implies error estimates in the observable expectations\REV{, which improve \cite[Theorem 7.4]{LasserLubich2020} from additive to uniform.}
\begin{corollary}\label{cor:expect}
Let the observable $O = \op_h(o)$ be the Weyl quantization of a symbol $o\in S(\RR^{2d})$. Let $t=ns$ and $0<h\leq 1$. Then the error of the observable expectations using either the first-order or second-order Trotter formulae is bounded by
\begin{align*}
    \lvert \bra \psi T_{1,n}(s) \ket \psi - \bra \psi T(t) \ket \psi  \rvert & \leq C_t s,
    \\
    \lvert \bra \psi T_{2,n}(s) \ket \psi - \bra \psi T(t) \ket \psi \rvert & \leq C_ts^2,
\end{align*}
for any wave function $\norm{\psi}_{L^2} = 1$, where $C_t$ is some constant independent of $n$ and $h$.
\end{corollary}
\begin{proof} By Cauchy-Schwarz inequality, we have
\begin{align*}
    &\lvert \bra \psi T_{1,n}(s) \ket \psi - \bra \psi T(t) \ket \psi  \rvert
     = \lvert \bra \psi T_{1,n}(s) - T(t)\ket \psi  \rvert
    \\ 
     \leq & \norm{\psi}_{L^2} \norm{\left(T_{1,n}(s) - T(t) \right) \psi}_{L^2} 
     \leq \|T_{1,n}(s)-T(t)\|_{L^2\to L^2}\norm{\psi}_{L^2}^2,
\end{align*}
which yields the desired result. The error bound for the second-order Trotter follows the same strategy.
\end{proof}

\section{Spatially discretized cases and discrete microlocal analysis}\label{sec:err_dis}
In this section, we first discuss the finite difference and pseudo-spectral types of spatial discretization, and then introduce the discrete microlocal analysis and how the resulting matrices from spatial discretization correspond to the discrete Weyl quantization of certain symbols. We establish the observable error bounds in the discrete case and discuss the number of Trotter steps.

\subsection{Trotter formulae in the spatially discretized setting}
For notational simplicity, we discuss the one-dimensional case here unless otherwise stated. High dimension follows the same argument, which will be mentioned at the end of \cref{sec:fd_dis} and \cref{sec:sp_dis}. Consider \cref{eq:semi_schd} on the computational domain $[a, b]$ with periodic boundary conditions. Note that the use of periodic boundary conditions is referred to as periodization that is commonly applied also in the case where the spatial domain of interests is the whole real line $\mathbb{R}$ but the wave function is expected to be negligible outside an interval $[a,b]$.

\subsubsection{Finite difference discretization}\label{sec:fd_dis}

We discretize the semiclassical Schr\"odinger \cref{eq:semi_schd} via a central finite difference spatial discretization with $N$ equidistant nodes $x_j = a +  (b-a)j/N$ with $0 \leq j \leq N-1$,
and have
$\I \partial_t \vec{\psi}(t) = (A + B) \vec{\psi}(t)$,
where $\vec{\psi}(t)$ is an approximation of the exact wave function $u$ of \eqref{eq:semi_schd} with its $j$-th entry as the approximated $u$ evaluated at $t$ and $x_{j-1}$, and with slight abuse of notation $A$ and $B$ are the spatial discretization of $- \frac{h^2}{2} \Delta$ and $V(x)$, respectively. We have
\begin{equation}\label{eq:A_fd}
    A = A^0_{\mathrm{fd}} :=  \frac{h^2 N^2}{2(b-a)^2} \left(\begin{array}{ccccc}
        2 & -1 & & & -1 \\
         -1& 2 & -1 & & \\
          & \ddots& \ddots& \ddots& \\
           & & -1& 2 & -1\\
        -1& & & -1 & 2 \\
    \end{array}\right)_{N \times N},
\end{equation}
and 
\begin{equation}\label{eq:B_fd}
    B = \text{diag}\left(V(x_0),V(x_1),\cdots,V(x_{N-1}) \right).
\end{equation}
Note that $A$ is a circulant matrix and hence can be diagonalized by a discrete Fourier transform, that is,
\begin{equation*} 
     A = \mathcal{F}_N^{-1} D_{\rm fd} \mathcal{F}_N,
\end{equation*}
where $\mathcal{F}_N$ is the discrete Fourier transform with $N$ modes and $D_{\rm fd} $ is a diagonal matrix with diagonal terms being the discrete Fourier transform of the first column of $A$. The discrete Fourier transform $\mathcal{F}_N :  \mathbb{C}^N \to \mathbb{C}^N$ is defined as
\begin{equation}\label{eq:dft}
(\mathcal{F}_N v)_k : = \hat v_k = \sum_{j = 0}^{N-1} v_j e^{-\I 2 \pi k \frac{x_j - a}{b-a}  } 
= \sum_{j = 0}^{N-1} v_j e^{-\I 2 \pi k j/N}, 
\end{equation}
for vectors $v$ of length $N$, and the inverse discrete Fourier transform $\mathcal{F}_N^{-1} :  \mathbb{C}^N \to \mathbb{C}^N$ as
\[
(\mathcal{F}_N^{-1} \hat v)_j : = \frac{1}{N} \sum_{k = 0}^{N-1} \hat v_k e^{\I 2 \pi k  \frac{x_j - a}{b-a} } 
= \frac{1}{N} \sum_{k = 0}^{N-1} v_k e^{\I 2 \pi k j/N}.
\]
Note that these are the definitions of the discrete Fourier transform and inverse discrete Fourier transform in numerical analysis, and the quantum Fourier transform corresponds to the unitary inverse discrete Fourier transform and vice versa.
When one combines this spatial discretization with the Trotter formulae temporally, the unitaries in the Trotter formulae become
\begin{equation} \label{eq:unitary_in_trotter_fd}
    e^{-\I As/h} = \mathcal{F}_N^{-1} e^{-\I D_{\rm fd}s/h} \mathcal{F}_N,
    \quad 
     e^{-\I Bs/h} = \text{diag}(e^{-\I s V(0)/h},\cdots, e^{-\I s V((N-1)/N)/h})
\end{equation}
for any time $s$. They are either diagonal in the physical space or diagonal in the Fourier basis and hence easy to implement. So is in higher dimensions, where the discretization becomes
\begin{equation}\label{eq:A_fd_hd}
A= A^0_{\mathrm{fd}}\otimes I_N^{\otimes d-1}+\dots +I_N^{\otimes d-1}\otimes A^0_{\mathrm{fd}}.
\end{equation}
The detailed error analysis for the finite difference discretization can be found in \cite{AnFangLin2021} and $N$ is chosen as $\Or(h^{-1})$. \REVV{As is discussed in the introduction, the choice of $N$ as $\Or(h^{-1})$ is standard in the literature, which results from the scaling of spatial discretization together with the standard approximation theory. See, e.g., \cite[Equation (4.2]{BaoJinMarkowich2002}, \cite[Equation (7.23) and the subsequent discussion]{LasserLubich2020}, \cite{Lubich2008book} and \cite{Singh2018thesis} for further details and justification. Note that under this choice of $N$ the matrix norm of $A$ is in fact bounded in $N$, in contrast to the non-semiclassical regime where $h=1$ considered in \cite{JahnkeLubich2000,HochbruckLubich2003} and so on.
}

\subsubsection{Pseudo-spectral discretization}\label{sec:sp_dis}

The idea of the Fourier pseudo-spectral discretization (or the Fourier collocation method) is to represent the wave function $u(t,x)$ as an expansion of trigonometric polynomials at every time $t$
\begin{equation} \label{eqn:w_trig_expansion}
    u(t,x) \approx u_N(t,x) = \sum_{k = - N/2}^{N/2-1} c_k(t) e^{\I \frac{2 \pi k}{b-a} (x -a)}, 
\end{equation}
where $k$ is the Fourier frequency and $x \in [a,b]$.
The coefficients $c_k$ are thus determined by letting $u_N(t,x)$ satisfy \eqref{eq:semi_schd} exactly at a number of collocation points $x_j = a + (b-a)j/N$, where $j = 0, \cdots, N-1$. One has
\begin{equation} \label{eqn:ham_schd_collocation}
    \I h\partial_t u_N(t,x_j)= -\frac{h^2}{2}  \Delta u_N(t,x_j)  + V(x_j) u_N(t,x_j),
\end{equation}
\begin{equation*}
    u_N(t, x_j)=  \sum_{k = - N/2}^{N/2-1} c_k(t) e^{\I 2\pi k j/N}.
\end{equation*}
Following the standard Fourier collocation calculations, the coefficient $\vec{c} = (c_k)$ satisfies
\begin{equation*} 
    \I h\partial_t  \vec{c} = D_{\rm sp} \vec{c}  + \mathcal{F}_N V_N \mathcal{F}_N^{-1} \vec{c}. 
\end{equation*}
Applying the inverse Fourier transform on both sides,
we obtain a system of ordinary differential equations of $\vec{\psi}(t) = (\psi_j(t))$ with $\psi_j(t) = u_N(t, x_j)$,
\begin{equation}  \label{eqn:schd_discrete_main}
    \I h\partial_t \vec{\psi} = \mathcal{F}_N^{-1} D_{\rm sp} \mathcal{F}_N \vec{\psi} + V_N \vec{\psi} : = (A  + B ) \vec{\psi},
\end{equation}
where $V_N := B$ is the same as defined in \eqref{eq:B_fd} and
\begin{equation*}
D_{\rm sp} = D_{\rm sp}^0 : = \frac{h^2}{2}\left(\frac{2 \pi}{b-a}\right)^2 \diag\left(   \left(- \frac{N}{2}\right)^2, \left(- \frac{N}{2} + 1\right)^2, \cdots,  \left(\frac{N}{2}-1\right)^2 \right).
\end{equation*}
The pseudo-spectral discretization is standard for this problem, and the discretization error can be found in, e.g., \cite{BaoJinMarkowich2002,LasserLubich2020,ChildsLengEtAl2022} and one needs $N = \Or(h^{-1})$. So is in higher dimensions, where the discretization becomes
\begin{equation}
D_{\mathrm{sp}}= D_{\mathrm{sp}}^0\otimes I_N^{\otimes d-1}+\dots +I_N^{\otimes d-1}\otimes D_{\mathrm{sp}}^0.
\end{equation}
When combined with the Trotter formulae temporally, the unitaries in the Trotter formulae are either diagonal in the physical space or the Fourier basis and hence can be efficiently implemented (see \cite{Su2021} for a perspective on fast-forwardable Hamiltonians).

\subsection{Semiclassical analysis on finite-dimensional spaces}\label{sec:discrete_microlocal} 

We now discuss the quantized $2d$-dimensional torus, which will allow us to rigorously analyze the observable error when spatial discretization is used. In particular, the Hilbert spaces considered are finite-dimensional, and pseudodifferential operators can frequently be set up to be discretizations of standard continuous operators. 

Discrete microlocal analysis takes place on an $N$-dependent Hilbert space called $H_N$, with operators $\op_N(a):H_N\to H_N$ playing the role of pseudodifferential operators. Such spaces and operators are easiest to define in a technical manner which directly relates them to those discussed in Section \ref{ss:microlocal}. For the reader new to microlocal analysis, however, \emph{it bears repeating that the spaces $H_N$ are finite-dimensional with a canonical orthonormal basis, and therefore can be viewed as equivalent to $\RR^{N^d}$, with the standard $\ell^2$ norm.} In this context, \emph{the operators $\op_N(a)$ are simply $N^d$-dimensional matrices, whose norm on $H_N$ is the standard spectral norm for matrices.}

We now begin with the definitions. Let $h=\frac{1}{2\pi N}$ for $N\in\NN$, and define the $N^d$-dimensional space \begin{equation}\label{eq:H_n}H_N:=\mathrm{span}\left\{N^{-\frac{d}{2}}\sum_{k\in\ZZ^d}\delta_{x=k+\frac{n}{N}}:n\in\{0,1,\dots,N-1\}^d\right\},\end{equation} where $\delta_{x=y}$ is the Dirac distribution at $y$. It can be shown that elements of $H_N$ are precisely the distributions on $\RR^d$ such that both they and their semiclassical Fourier transforms $\mathcal{F}_hu$ (see (\ref{eq:scft})) are periodic with period $1$. For notational convenience, define $$Q_n=N^{-\frac{d}{2}}\sum_{k\in\ZZ^d}\delta_{x=k+\frac{n}{N}}.$$ We decide by definition to have $(Q_n)$ be an orthonormal basis of $H_N$, which endows $H_N$ with a Hilbert space structure. Henceforth, we may represent operators on $H_N$ as $N^d$ by $N^d$ matrices in the basis $(Q_n)$. Such a representation is made especially natural by the fact that the semiclassical Fourier transform $\mathcal{F}_h$ acts as in the $(Q_n)$ basis as the discrete Fourier transform matrix $\mathcal{F}_N$.

For those interested in geometric quantization, we also remark that the construction of $H_N$ is a specific example of the general geometric Toeplitz quantization studied by e.g. Deleporte \cite{Deleporte2019}. We now give an overview of the pseudodifferential calculus on $H_N$, which is outlined more extensively in the works by Christiansen–Zworski \cite{ChristiansenZworski2010}, Schenck \cite{Schenck2009}, and (in slightly more generality) Dyatlov–Jezequel \cite{DyatlovJezequel2021}, and which will suffice to discuss the spatial discretization in our context. Let $a\in \cinf(\TT^2)$. Then $a$ lifts to a doubly-periodic function $\tilde{a}$ on $T^*\RR^d$, which belongs to the symbol class $S$. Define \begin{equation}\label{eq:H_N_quant_def}\op_N(a):=\op_h(\tilde{a})|_{H_N}.\end{equation} This is a map from $H_N$ to itself that may also be given in coordinates by $$\op_N(a)Q_j=\sum_{m=0}^{N-1}A_{mj}Q_m$$ with \begin{equation}\label{eq:coords}A_{mj}=\sum_{k,l\in\ZZ^d}\hat{a}(k,j-m-lN)(-1)^{\la k,l\ra}e^{\pi \I\frac{\la j+m,k\ra}{N}
}.\end{equation} 

Most results from standard pseudodifferential calculus carry over to $H_N$, see \cite{ChristiansenZworski2010},
\cite{DyatlovJezequel2021} for details. In particular, pseudodifferential operators on $H_N$ are bounded independently of $N$, with the bound given by the norm of their analogue on $L^2(\R^d)$.
\begin{lemma}\label{lem:H_N_L^2_bound}
Let $a\in\cinf(\TT^{2d})$, and let $\tilde a$ be the lift of $a$ to a periodic function on $T^*\RR^d$. Then
\begin{equation}\label{eq:H_N_norm_bound}
\|\op_N(a)\|_{H_N\to H_N}\le\|\op_h(\tilde{a})\|_{L^2(\RR^d)\to L^2(\RR^d)}.
\end{equation}
\end{lemma}
We remark once again that the space $H_N$ is isomorphic to $\RR^{N^d}$, so the left side of (\ref{eq:H_N_norm_bound}) is nothing more than $\|\op_N(a)\|_{\ell^2\to\ell^2}$ when $\op_N(a)$ is treated as a matrix. Results like Lemma \ref{lem:H_N_L^2_bound} can be found in several sources, including \cite{BuozouinaDeBievre1996,DyatlovJezequel2021}, for various settings.
We provide a proof sketch in the supplementary materials~\cite{thesupplement} for our scenario.
Finally, we also mention that one may bound $\|\op_N(a)\|_{H_N\to H_N}$ directly in terms of the derivatives of $a$ in a more elementary manner, see \cite[Proposition 2.7]{ChristiansenZworski2010}.

We now pause to give examples of pseudodifferential operators on $H_N$, which include some very natural operators. Suppose that $d=1$, so $\dim H_N=N$. Then in the case that $a(x,\xi)=V(x)$ is independent of $\xi$, we have $\op_N(a)Q_n=a(n/N)Q_n$. In other words, as a matrix $$\op_N(a)=\diag(V(0),V(1/N),\dots, a((N-1)/N)),$$ which is the matrix $B$ given in (\ref{eq:B_fd}) defined on the torus $[0,1]$. Similarly, if $b(x,\xi)=W(\xi)$ is independent of $x$ then $$\op_N(b)=\mathcal{F}_N^{-1}D\mathcal{F}_N,$$ where $\mathcal{F}_N$ is the discrete Fourier transform matrix and $D=\diag(W(0),W(1/N),\dots,$ $ W((N-1)/N))$. Specifically, $\op_N(b)$ is a circulant matrix.
In particular, $A$ in the finite difference discretization as defined in \eqref{eq:A_fd} is (up to a constant factor) the quantization of $b(x,\xi)=2-2\cos(2\pi\xi)$ on the torus.
When $d>1$, we analogously get (\ref{eq:A_fd_hd}) as  $\op_N\left(\sum_{j=1}^d(1-\cos\xi_j)\right).$

In the following discrete analysis, we mainly focus on the finite difference spatial discretization, as it is the natural outcome for discrete Weyl quantization on a torus. Our result also works for the FFT pseudo-spectral discretization, with a modifier applying to the kinetic part, which is to be detailed in \cref{rem:modifier}. 

By applying Theorems \ref{thm:long_trot1} and \ref{thm:long_trot2} to $\tilde{a},\tilde{b},$ and $\tilde{o}$ and invoking Lemma \ref{lem:H_N_L^2_bound}, we have the following theorems on the Trotter observable errors in the spatially discrete case. The norms below are the standard spectral (i.e., $\ell^2\to\ell^2$) norms on $\RR^{N^d}$, which are once again identical to the operator norms on $H_N$ due to the identification made via the orthonormal basis $(Q_n)$, see Section \ref{sec:discrete_microlocal}.

\begin{theorem}\label{cor:disc_trot1}
Let $a,b,o\in\cinf(\TT^{2d})$ and $A,B,O:H_N\to H_N$ be their respective quantizations, as defined by (\ref{eq:H_N_quant_def}), and let $t=ns$ and $0<h \leq 1$. Denote the time evolution under the exact dynamics as
\begin{equation}\label{eq:ob_exact}
    T(t): = e^{\I Ht/h} O e^{-\I Ht/h},
\end{equation}
and under the first-order Trotter formula as
\begin{equation}\label{eq:ob_exact}
    T_{1,n}(s): = (e^{\I As/h} e^{\I Bs/h})^n O (e^{-\I Bs/h} e^{-\I As/h})^n. 
\end{equation}
The difference between the two can be estimated as
\begin{equation}\label{eq:ob_err_trotter1}
    \norm{ T_{1,n}(s) - T(t)} \leq C_t s.
\end{equation}
\end{theorem}

\begin{theorem}\label{cor:disc_trot2}
Let $a,b,o\in\cinf(\TT^{2d})$ and $A,B,O:H_N\to H_N$ be their respective quantizations, as defined by (\ref{eq:H_N_quant_def}), and let $t=ns$ and $0<h \leq 1$. Denote the time evolution under the exact dynamics as
\begin{equation}\label{eq:ob_exact}
    T(t): = e^{\I Ht/h} O e^{-\I Ht/h},
\end{equation}
and under the second-order Trotter formula as
\begin{equation}\label{eq:ob_exact}
    T_{2,n}(s): = (e^{\I As/(2h)} e^{\I Bs/h}e^{\I As/(2h)})^n O (e^{-\I As/(2h)}e^{-\I Bs/h} e^{-\I As/(2h)})^n. 
\end{equation}
The difference between the two can be estimated as
\begin{equation}\label{eq:ob_err_trotter2}
    \norm{ T_{2,n}(s) - T(t)} \leq C_t s^2 .
\end{equation}
\end{theorem}

We have the following estimate for the number of the Trotter steps, which is also the number of queries to $\exp(-\I At/h)$ and $\exp(-\I Bt/h)$.
\begin{corollary}[Query complexity] \label{cor:query}
We use the central finite difference for spatial discretization and Trotter formulae for time discretization to obtain an $\epsilon$-approximation in the operator norm of the time-evolved observables satisfying the assumptions of \cref{cor:disc_trot1} and following the semiclassical Schr\"odinger equation \eqref{eq:semi_schd}. 
    Let $L_{1}$ and $L_{2}$ denote the total number of required time steps of first-order and second-order Trotter formulae, respectively. Then for sufficiently small $\epsilon$ and $\Or(1)$ evolution time $t$, we have
\begin{equation}
    L_1 = \Or(1/\epsilon), \quad L_2 = \Or(1/\epsilon^{1/2}),
\end{equation}
independent of the Planck constant $h$.
\end{corollary}

\begin{remark}\label{rem:modifier}
More generally, letting $\phi\in\cinf(\TT^1)$, we may let $$A=\frac{1}{(2\pi)^2}\op_N\left(\sum_{j=1}^d\phi(\xi_j)\right).$$ If $\phi$ is nearly quadratic in the momentum region of interest, this operator corresponds to an ``approximate spectral discretization" of the Laplacian. For instance, one may take $\phi(\xi)=\xi^2\chi(\xi)$, where $\chi\in C^{\infty}_0\left(-\frac{1-c}{2},\frac{1-c}{2}\right)$ and $\chi(\xi)=1$ on $\left[-\frac{1}{2}+c,\frac{1}{2}-c\right]$.

The reason to introduce such a modifier is that the FFT-discretized kinetic operator $\widehat{p}^2/2 = \op_N(\xi^2)$ is the discrete Weyl quantization of the function $\xi^2$, which is not necessarily periodic on the computational domain $\mathcal{D}$. When $\mathcal{D} = [-\pi, \pi]$, though the function values become periodic, its derivative is discontinuous on the boundary. But note that the computational domain is chosen so that the solution remains supported with in it. In other words, though the function $\xi^2$ can have ill-defined high-order derivatives on the boundary, the solution almost vanishes there and hence the influence of the boundary effects is negligible. Applying the modifier takes into account of this simulation feature. In this case, we can have $\phi(\xi) \in C^\infty$ so that our analysis applies and the query complexity in \cref{cor:query} holds.
\end{remark}

\section{Numerical Results} \label{sec:numerics}

In this section, we demonstrate the numerical results of the observable error bounds for the semiclassical Schr\"odinger equation simulation. 
For simplicity, we consider the following Hamiltonian
\begin{equation}
H= \REV{-\frac{h^2}{2}\Delta + V(x) }, \quad V(x) = \cos(x), \quad x\in [-\pi,\pi]
\label{eqn:ham_lap_v}
\end{equation}
with periodic boundary conditions. Here $A$ corresponds to the discretized $-\Delta$ using the second order finite difference scheme, and $B$ the discretized $V(x)$. The number of spatial grids is fixed as $N = 1/h$. Though our theoretical results hold as long as $N = \Or(1/h)$, the particular choice of $N$ as $1/h$ instead of $1/(2\pi h)$ is because the domain here is $[-\pi, \pi]$, instead of $[-1,1]$.
We illustrate our theoretical results using the following two smooth observables: the cosine observable as the quantization of $o(x,p) = \cos(x)$; and the momentum observable $\hat p = -\I h \nabla_x$. The errors are measured in the operator norm between the observable evolution matrices.

\begin{figure}
    \centering
    \includegraphics[width=.45\textwidth]{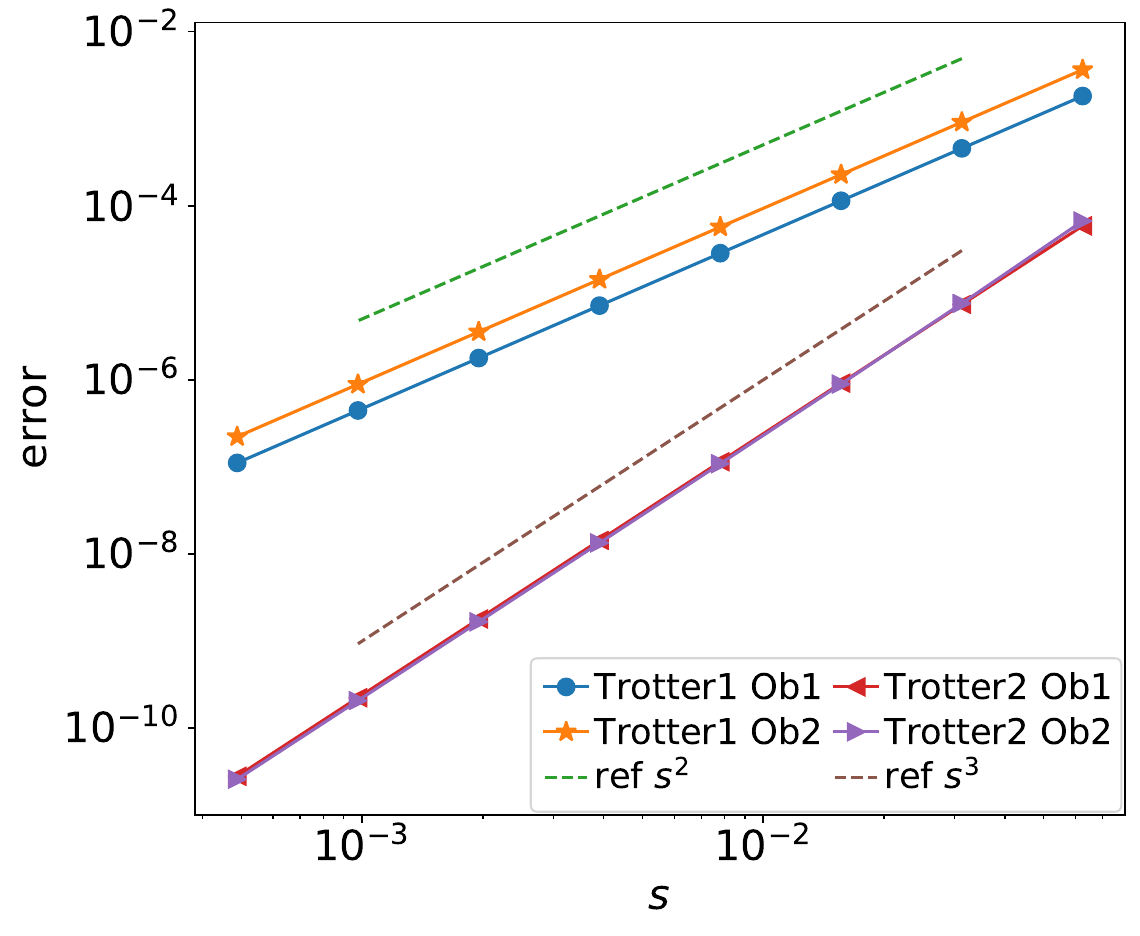}
    \caption{Log-log plot of the scaling of the operator norm of the difference of the observable evolution matrices $T_\ell(s) - T(s)$ (for $\ell = 1,2$) for various time step size $s$. ``Trotter1" labels the first-order Trotter formula while ``Trotter2" labels the second-order Trotter formula. ``Ob1" is for the cosine observable that is the quantization of $\cos(x)$ and ``Ob2" denotes the momentum observable $\hat p = -\I h \nabla_x$. The reference line is for asymptotic scaling in $s$. }
    \label{fig:lte_s}
\end{figure}

First, we verify the convergence rate with respect to the time step size $s$ for the short-time evolution numerically for this periodic Hamiltonian. The Planck constant $h$ is fixed as $1/64$ and the time step $s = 2^{-4}, 2^{-5}, \cdots, 2^{-11}$. The system is simulated for a single time step using both the first-order and the second-order Trotter formulae, and the number of spatial discretization $N$ is fixed as $1/h$. \cref{fig:lte_s} plots the operator norm of $T_\ell(s) - T(s)$ (for $\ell = 1,2$) with the finite difference discretization in the log-log scale. It can be seen that the short-time error of the first-order Trotter formula scales quadratically with respect to the time step size $s$ while the second-order Trotter formula scales cubically in $s$, which agrees with \cref{cor:error_short_time}.

\begin{figure}
    \centering
    \includegraphics[width=.45\textwidth]{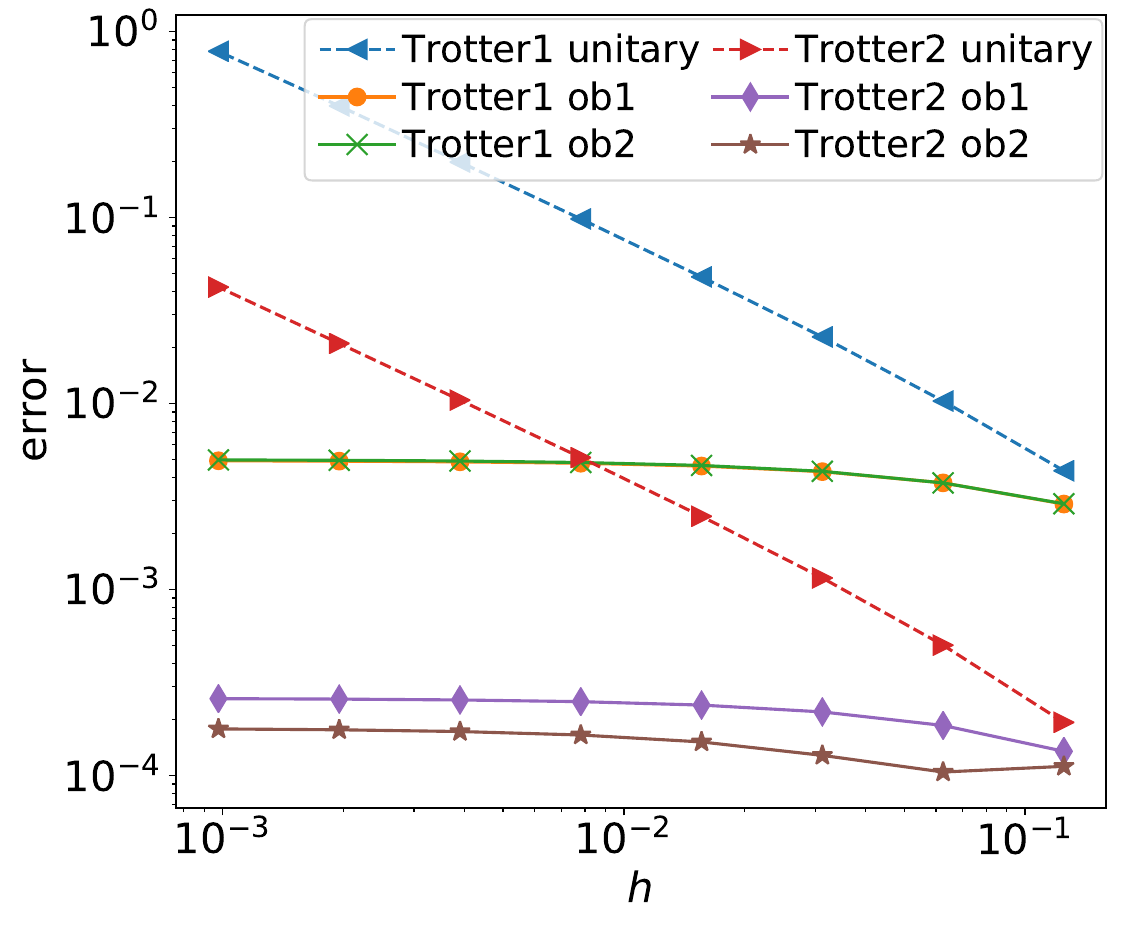}
    \caption{Log-log plot of the errors in the operator norm for various Planck constants $h$. ``unitary" denotes the error measuring in the operator norm of the unitaries, while ``ob1" and ``ob2" measures the operator norm error of the observable evolution matrices. The errors in the unitary scales as $1/h$. However, observable errors do not grow as $h$ decreases. }
    \label{fig:lte_h}
\end{figure}

We then present the short-time error scaling in terms of the Planck constant $h$. \cref{fig:lte_h} plots the operator norm error versus $h$ in the log-log scale. Here the values of $h$ are chosen as $2^{-3}, 2^{-4}, \cdots, 2^{-10}$, and the time step size $s$ is fixed to be $0.1$. Note that this time step size is relatively large (compared to $\Or(h)$). It can be seen that for both first-order and second-order Trotter formulae, the operator norm error in terms of the unitaries increases as $h$ decreases. In fact, the scaling is approximately $1/h$. However, the operator norm errors in terms of the observables do not grow as $h$ decreases, as is proved rigorously in \cref{cor:error_short_time}.

\begin{figure}
    \centering
    \subfloat[for various $s$]{
        \includegraphics[width=.4\textwidth]{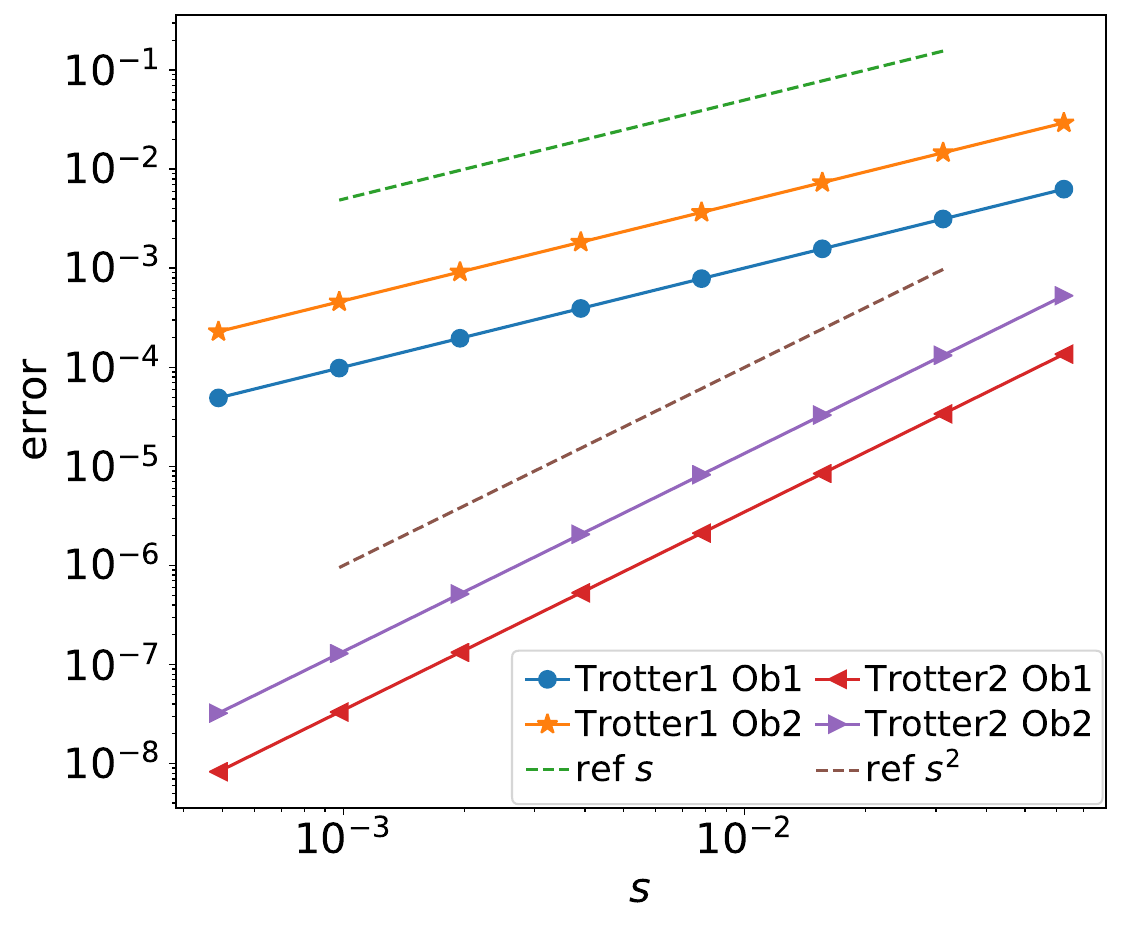} \label{fig:long_s}}
    \subfloat[for various $h$]{
        \includegraphics[width=.4\textwidth]{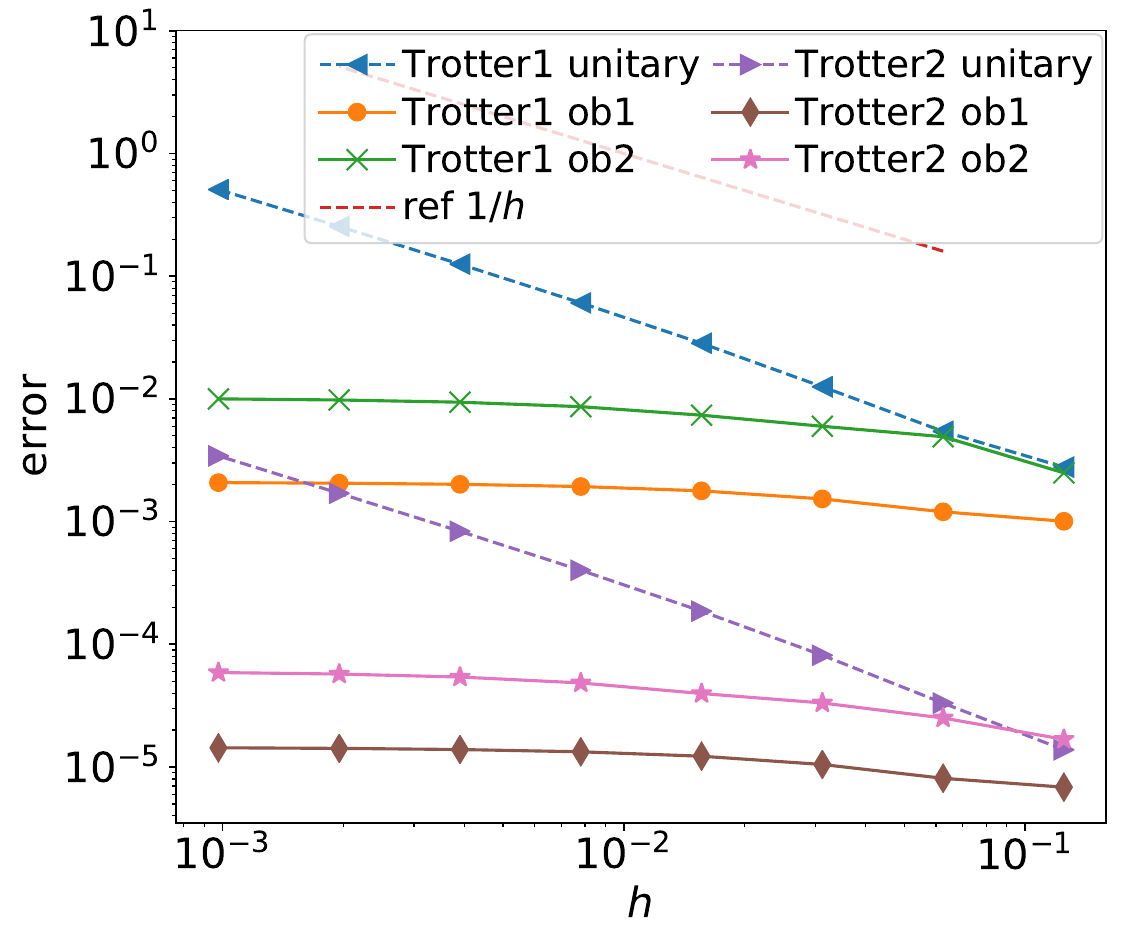} \label{fig:long_h}}
    \caption{Long-time error in the operator norm of the unitaries and the observable for various $s$ (Left) or $h$ (Right). The final time is $1$. ``unitary" denotes the error measuring in the operator norm of the unitaries, while ``ob1" and ``ob2" measure the operator norm error of the observable evolution matrices. For both Trotter formulae, the long-time errors are uniform in $h$. The first-order Trotter formula has a long-time observable error bound of $\Or(s)$ and the second-order one exhibits $\Or(s^2)$. }
    \label{fig:long}
\end{figure}

We now demonstrate the long-time error in both the $s$ and $h$ scalings. The dynamics is simulated till the final time $t = 1$. For the scaling in $s$, we fix $h$ as $1/256$ and chose the time step size $s$ as $2^{-4}$, $2^{-5}$, $\cdots$, $2^{-11}$. \cref{fig:long_s} plots the error in the operator norm of the observables in log-log scale. As proved in \cref{thm:long_trot1} and \cref{thm:long_trot2}, the long-time errors for the first-order Trotter formula are $\Or(s)$ while those for the second-order Trotter formula are $\Or(s^2)$. 
Next, we verify the uniformity of the error bounds in $h$. The time step size $s$ is fixed to be $0.02$.
The Planck constants $h$ are chosen as $2^{-3}, 2^{-4}, \cdots, 2^{-10}$ and $N = 1/h$. The operator norm error of the unitaries and the observables are plotted in \cref{fig:long_h} in the log-log scale. It can be seen that the operator norm error in terms of the unitaries for both the first-order and second-order Trotter formulae increases as $h$ decreases, but the observable errors remain uniformly bounded in terms of $h$. Both figures verify the error bounds proved in \cref{thm:long_trot1} and \cref{thm:long_trot2}.

\section{Conclusion and discussion} \label{sec:conclusion}

We have studied the observable error bounds, i.e.\ the error of the time-evolved observables measuring in their operator norm, of both the first and second-order Trotter formulae for the semiclassical Schr\"odinger equation, and have proven uniform-in-$h$ observable error bounds without sacrificing the order of convergence. This results in an estimate of the number of Trotter steps that is independent of the small parameter $h \ll 1$, while state-of-the-art error bounds measuring in the operator norm can have a polynomial overhead in terms of $h^{-1}$. 
Note that it is well known that complexity improvements can be made when considering the observable error bounds for Hamiltonian with geometric locality properties, which can be taken advantage of to get bounds of the Lieb-Robinson type. It is interesting to note that the Sch\"odinger equation processes a non-local unbounded operator. Here we demonstrated that there can be merits to study the observable error bounds even when the underlying Hamiltonian is nonlocal, and there can exist other mechanisms (than locality) to gain such observable error improvement. As a specific example of the dynamics involving multi-scales, we demonstrated that the existence of the small parameter in a multiscale problem can sometimes be taken advantage of to improve the complexity of its algorithms. 

In terms of analysis, our results go beyond the existing literature in the following sense: First, our error bound is uniform in $h$ and without reducing the convergence order of the algorithm. This is, to our knowledge, the first uniform-in-$h$ observable error bound for the \REV{linear} semiclassical equation reproducing the same order of convergence as the numerical schemes. It is worth noting that this is done by a direct estimate of the error between the Trotter formulae and the Schr\"odinger equation, instead of passing the limit in terms of the small parameter as in the typical proof strategy of such problem which can inevitably make the error bounds additive. Moreover, we extend our continuous-in-space results to the setting with spatial discretization directly using ``discrete microlocal analysis" on the quantized torus, where we take advantage of the correspondence of matrices and the Weyl quantization on a torus. This allows for operators on finite-dimensional Hilbert spaces, such as those entered into quantum or classical computation devices, to be considered on their own terms, and closes the gap in the observable error bound studies with the spatial discretization. 
To our knowledge, in the setting of quantum simulation this is the first analysis of discrete operators using microlocal techniques.

Our result focuses on symbol-classed observables that are quantization of $C^\infty$ functions.
However, it is worth noting that it can still hold when the quantization is not smooth on the boundary, such as polynomials of the momentum operators $p$ on a torus that can have ill-defined derivative, provided the wave function is negligible near the boundary. It is also interesting to consider the matrices that correspond to quantization of non-smooth functions. In that case, one needs to use the regularity of the solution. Though an operator norm bound of the time-evolved observable is not hopeful, the observable expectation value result (like \cref{cor:expect}) is possible for smooth initial conditions, which serves as an interesting future direction. As a separate matter, though we focus on the case of $t = \Or(1)$, the actual time scaling that it corresponds to is $\tilde{t} = t/h \gg 1$. It is an interesting future direction to work out the explicit dependence on $t$, which likely reflects the classical ``chaos" of the system.

There is also further work to be done with new algorithms in the setting of this paper. One possibility is to derive analogous bounds for higher-order Trotter methods, as discussed in \cite{DescombesThalhammer2010,ChildsSuTranEtAl2020}.
We expect that the higher-order analogs of Theorems \ref{thm:long_trot1} and \ref{thm:long_trot2} hold, providing $O(s^k)$ error for $k$-th order Trotter. Additionally, one could study observable error for other splitting algorithms or exponential integrators, such as \cite{Singh2015,BaderIserlesKropielnickaSingh2014,BlanesCasas2021}. 
Finally, we would like to remark that though the semiclassical Sch\"odinger equation can be derived from the molecular dynamics under the Born-Oppenheimer approximation, it still requires one to compute the potential $V(x)$ in \eqref{eq:semi_schd} from the electronic structure which is a non-trivial task. Though serving as a natural and interesting future direction, a generalization to the full nucleus-electron dynamics is non-trivial for at least two reasons: first, the total potential contains the Coulomb interaction which does not satisfy the regularity assumption though the electronic potential energy surfaces can; furthermore, for full nucleus-electron molecular dynamics, more than one potential energy surface is typically involved and there can be non-adiabatic phenomena where nuclei tunnel between different surfaces and the Born-Oppenheimer approximation breaks down. The investigation of the full molecular Hamiltonian near Born-Oppenheimer is a promising direction left for future study.

\section*{Acknowledgments:} 

This work was partially supported by NSF grant number DMS-2208416 (D.F.), by NSF grant DMS-1952939 (Y.B.). D.F. also acknowledges the support of Ralph E. Powe Junior Faculty Enhancement Award, the support from the NSF QLCI program and the Challenge Institute for Quantum Computation funded by NSF through grant number OMA-2016245, and the hospitality of the Kavli Institute for Theoretical Physics funded by NSF under Grant No. NSF PHY-1748958. 
The authors thank Lin Lin and Maciej Zworski for discussions.

\bibliographystyle{unsrt}
\bibliography{ref}

\appendix
\section{Derivation of Table 1} \label{app:derivation_table}
We recall that the spatial discretization of the semiclassical Hamiltonian $H^h = H/h$ is denoted
as $A^h + B^h$, where $A^h$ and $B^h$ are the discretized matrices of $- \frac{h}{2} \Delta$ and $\frac{1}{h} V(x)$, respectively. We see that
\begin{equation*}
    \norm{A^h} = \Or(h^{-1}), \quad \norm{B^h} = \Or(h^{-1}),
\end{equation*}
\begin{equation*}
    \norm{[A^h,B^h]} = \Or(h^{-1}), \quad \norm{[A^h, [A^h,B^h]]} = \Or(h^{-1}), \quad \norm{[B^h, [A^h,B^h]]} = \Or(h^{-1}).
\end{equation*}

\subsection{$p$-th order Trotter formulae}
It has been shown that the $p$-th order Trotter formulae has nested commutator scaling that depends on the $(p+1)$-th order commutator~\cite{ChildsSuTranEtAl2020}
\begin{equation} \label{eq:comm_p_trotter}
\sum_{\ell_1, \cdots, \ell_{p+1} = 1}^2\norm{[H_{\ell_{p+1}}, [H_{\ell_p}, \cdots [H_{\ell_2}, H_{\ell_1}]]]},
\end{equation}
where $H_1 = A^h$ and $H_2 = B^h$ are the discretizations of $- h \Delta/2 $ and $V/h$. In order to make clear the cancellation in the nested commutators and estimate the norm of them, we need some algebra preparation. Note that $\Delta = \nabla_x^2$ and $V$ in fact generate a free Lie algebra. For simplicity, we consider one-dimensional case.
 
Denote $\mathcal{F}$ as the free Lie algebra generated by $\partial_x^2$ and $V$. The key observation is that $\mathcal{F}$ can be embedded in a larger Lie algebra $\mathcal{L}$, where in the latter algebra commutators has simpler structures. Define the Lie algebra $\mathcal{L}$ as 
$$\mathcal{L}: = \left\{\sum_{k=0}^n y_k(x) \partial_x^k,\, n \in \mathbb{Z}_+,\, y_0, \cdots, y_n \text{ are smooth and periodic} \right\}.$$
We call $n$ the height (``ht'') of an element in $\mathcal{L}$. One can explicitly compute the terms to find its height. For example, 
\begin{align*}
[V, \partial_x^2] = -(\partial_x^2 V) - 2(\partial_xV)\partial_x
\end{align*}
is of height 1. Such cancellation exists in general and we have the following height reduction lemma, which can be found in, e.g., \cite[Corrolary 6]{Singh2015} (see also \cite{IserlesKropielnickaSingh2018,IserlesKropielnickaSingh2019}).

\begin{lemma}[Height Reduction in Lie Algebra] \label{lmm:ht_red}
$$ht([A,B]) \leq ht(A) + ht(B) -1, \quad \forall A,B \in \mathcal{L}.$$
\end{lemma}
This can be shown by straightforward calculations where the highest term is canceled. An immediate consequence of the lemma is that the highest height of the nested commutator with $(p+1)$-terms is $p$. In other words, the nested commutator of $p+1$ with the highest height 
\[
[\underbrace{\partial_x^2, \cdots [\partial_x^2}_{p \text{ times}}, V(x)]]
\]
is a $p$-th order differential operator and so is
\begin{equation}\label{eq:h_comm_p_1}
[\underbrace{-\frac{h}{2}\partial_x^2, \cdots [-\frac{h}{2} \partial_x^2}_{p \text{ times}}, \frac{1}{h}V(x)]] 
\end{equation}
with the prefactor of $h$ being $h^{p-1}$. Under the meshing strategy with the number of spatial grids being fixed as $\Or(h^{-1})$, we have the discretization of the $p$-th derivative being $\Or(h^{-p})$. Therefore, the spatial discretization of \eqref{eq:h_comm_p_1} has the operator norm of $\Or(h^{p-1} h^{-p}) = \Or(h^{-1})$. 

In fact, this is true for every term in the nested commutator \eqref{eq:comm_p_trotter}. Suppose a nested commutator with $p+1$ terms consists of $m$ terms of $-\frac{h}{2}\partial_x^2$ and $p+1-m$ terms of $\frac{1}{h}V(x)$. The prefactor of $h$ is $h^m h^{- p- 1 + m} = h^{-p-1+2m}$. On the other hand, according to \cref{lmm:ht_red}, the height (or the order of derivatives) is bounded by $2 m - p$. This is because, $\partial_x^2$ has height $2$ and is repeated $m$ times while there are $p$-layers of commutators so that we can apply order reduction $p$ times. Therefore, the spatial discretization of this nested commutator has the operator norm at most $\Or(h^{- 2m + p} h^{-p-1+2m}) = \Or(h^{-1})$. We can now conclude that  
\begin{equation} 
\alpha_{\rm comm} : = \sum_{\ell_1, \cdots, \ell_{p+1} = 1}^2\norm{[H_{\ell_{p+1}}, [H_{\ell_p}, \cdots [H_{\ell_2}, H_{\ell_1}]]]} = \Or(h^{-1}).
\end{equation}
Therefore, the number of Trotter steps of $p$-th order Trotter formulae to achieve precision $\epsilon$ to the unitary evolution operator time is
\begin{equation}\label{eq:query_p_trotter}
\Or\left( \frac{\alpha_{\rm comm}^{1/p} t^{1+1/p}}{\epsilon^{1/p}} \right) = \Or\left( \frac{ t^{1+1/p}}{h^{1/p} \epsilon^{1/p}} \right),
\end{equation}
with the number of one-dimensional spatial grids fixed as $\Or(h^{-1})$ and not changing with the simulation time $t$.
It is interesting to note that going to the vector norm error bound for the wave function, the scaling remains the same as \eqref{eq:query_p_trotter} for the semiclassical Hamiltonian (see \cite{DescombesThalhammer2010} and its reproduction in \cite{JinLiLiu2021}).

\subsection{Truncated Taylor series}
Consider the truncated Taylor approach for the semiclassical Hamiltonian $H^h : =  H/h$. One takes the number of time steps $L$ such that $L \geq \|H^h\|t = \|H\| t/ h$ and has
\[
\exp\left( \frac{-\I H^h t}{L}\right) = \exp\left( \frac{-\I H t}{ hL}\right)  \approx 
\sum_{k=0}^K \sum_{\ell_1, \cdots, \ell_k = 1}^2 \frac{(-it)^k}{(L h)^k k!} H_{\ell_1} \cdots H_{\ell_k},
\]
where $H_1 = A$ and $H_2 = B$.
According to \cite{BerryChildsCleveEtAl2014}, the overall complexity is given by the number of segments $L$ times $K$, where 
$$K = \Or{\left( \frac{\log(\frac{L}{\epsilon})}{\log\log(\frac{L}{ \epsilon})}\right)}.$$
Applying the LCU lemma \cite{ChildsWiebe2012},
the computational time is proportional to the absolute sum of coefficients
\[
\sum_{j \in J} \frac{(t)^k}{(L h)^k k!} = e^{t/(L h)},
\]
where $J: = \left\{(k, \ell_1, \cdots, \ell_k): k\in \mathbb{N}, \ell_1, \cdots, \ell_k \in \{1,2 \right\} \}$. Following the algorithm, one takes $t/(Lh) = \ln(2)$. So that the sum equals 2 and $L = \lceil t/(h \ln(2))\rceil$. Therefore the query complexity using truncated Taylor algorithm is
$$\Or\left( \frac{t}{h} \frac{\log(\frac{t}{h\epsilon})}{\log\log(\frac{t}{ h\epsilon})} \right).$$
Note that this also agrees with \cite[Theorem 2]{ChildsLengEtAl2022}, in which $\norm{f}_{\rm max} = \Or(h^{-1})$ for our case.

\subsection{QSVT}
The quantum signal processing approach~\cite{LowChuang2017} and its generalization of quantum singular value transformation (QSVT)~\cite{GilyenSuLowEtAl2019} achieve the optimal query complexity in all parameters for time-independent Hamiltonian simulation. For $\Or(1)$ time, the scaling in \cite[Theorem 3]{LowChuang2017} is 
$$\Or\left(d\|H\|_{\max} + \frac{\log(1/\epsilon)}{\log\log(1/\epsilon)} \right). $$
Here $d$ is the sparsity of the Hamiltonian, $\|H\|_{\max}$ denotes the largest matrix element of $H$ in absolute value. In our case, since $\|H\|_{\max} = \Or(h^{-1})$ the query complexity has an $h^{-1}$ scaling.

\subsection{Interaction Picture}
As is discussed in the introduction, the operator norm of the Hamiltonian in the interaction picture is
\[
\norm{H_I(t)} =  \norm{ e^{\I A t} B e^{-\I A t}} = \Or(h^{-1}).
\]
Therefore, by either the truncated Dyson series approach \cite[Theorem 7]{LowWiebe2019} or the more recent rescaled Dyson series approach \cite[Theore 3]{ChildsLengEtAl2022}, the query complexity scaling in $h$ is $\widetilde{\Or}({h^{-1}})$ in the semiclassical case.

 \section{QFT and circulant matrices} \label{app:circulant_circuit}

 In this section, we clarify the relationship of the diagonalization of a circulant matrix and quantum Fourier transform. 
 Given a $N\times N$ circulant matrix $M$, it can be diagonalized by a discrete Fourier transform of $N$ modes defined in \eqref{eq:dft}, denoted as $\mathcal{F}$.
 We have
 \begin{equation}
 M = \mathcal{F}^{-1}\diag(\mathcal{F}\vec{v}) \mathcal{F}.
 \end{equation}
 where $\vec{v}$ is the first column of the matrix $M$, and $D = (D_{kk})$ is a diagonal matrix with its diagonal terms being the discrete Fourier transform of the first column of $M$. 
 Note that the discrete Fourier transform and the quantum Fourier transform (denote as $Q$) admit the following relationship:
 \[
 Q = \sqrt{N} \mathcal{F}^{-1} = \frac{1}{\sqrt{N}} \mathcal{F}^{\dagger}, \quad Q^\dagger = \frac{1}{\sqrt{N}} \mathcal{F}.
 \]
 Therefore, $M$ can be diagonalized via quantum Fourier transform
 \[
 M = Q D Q^\dagger.
 \]
 In fact, in our case for the Trotter formulae, both $e^{-\I As/h}$ and $e^{-\I Bs/h}$ are unitaries, so that $\lvert D_{kk} \rvert = 1$ for all $k$. The circuit implementation is standard, see, e.g., \cite[Chapter 6]{Lin2022} and \cite[Section 6]{AnFangEtAl2022}.

\section{Sketch of Proof of Lemma 4.1}\label{app:H_N_bound}

For completeness, we provide a sketch of the proof of Lemma 4.1, which bounds the norm of the discrete operator $\op_N(a)$ with the operator norm of $\op_h(\tilde a)$ on $L^2(\RR^d)$. Some of the details can be found in \cite{BuozouinaDeBievre1996} (for $d=1$) and \cite{DyatlovJezequel2021} (for general $d$). We assume familiarity with the concept of a direct integral of Hilbert spaces, see \cite{ReedSimonIV1978,hall_book}.

We begin by defining a natural generalization of the space $H_N$. Once again set $h=\frac{1}{2\pi N}.$ Let $w=(k,\kappa)\in\RR^{2d}$, and consider the functions $$a_w(x,\xi):=e^{\frac{i}{h}(\la\kappa,x\ra-\la k, \xi\ra)}.$$ Disregarding for the moment that $a_w$ is not actually in $S$ (as its seminorms depend on $h$), we quantize it with anyway via Equation (2.1) of the article and define $U_w:=\op_h(a_w)$, with explicit formula $$U_wf(x)=e^{\frac{i}{h}(\la\kappa,x\ra-\la k, \xi\ra)}f(x-k)$$ which gives $U_w$ to be a unitary operator on $L^2(\RR^d)$. We remark that $U_w$ can be viewed as a ``phase space translation" by $w$, which is justified by the Egorov-type formula 
\begin{equation}\label{eq:phasetranslation}
U_w^{-1}\op_h(a)U_w=\op_h(b)\end{equation} where $b(x,\xi)=a(x+k,\xi+\kappa)$.

Let $\theta=(\theta_x,\theta_{\xi})\in\TT^{2d}$ be a parameter, and define the space $H_{N,\theta}\subseteq\mathcal{S}'(\RR^d)$ as $$H_{N,\theta}=\{u\in\mathcal{S}'(\RR^d):U_wu=e^{2\pi i(\la\theta_{\xi},k\ra-\la \theta_x,\kappa\ra)+N\pi i\la k,\kappa\ra}u\,\,\,\forall w=(k,\kappa)\in\ZZ^{2d}\}.$$

It is easy to check that $H_{N,\theta}$ is $N^d$-dimensional, and that $H_{N,0}=H_N$ as defined by (4.10) of the article. Moreover, $H_{N,\theta}$ has basis $(Q_{n,\theta})$ (which we again take to be orthonormal) with $$Q_{n,\theta}=N^{-\frac{d}{2}}\sum_{k\in\ZZ^d}e^{-2\pi i\la\theta_{\xi},k\ra}\delta_{x=k+\frac{n-\theta_x}{N}}.$$

A key result \cite[Lemmas 2.5 and 2.6]{DyatlovJezequel2021} says that $L^2(\RR^d)$ can be decomposed as a direct integral of the $H_{N,\theta}$'s. Specifically, for $f\in\mathcal{S}(\RR^d)$, let $$\Pi_{N,\theta}(f)=\sum_{n\in\{0,\dots N-1\}^d}\la f,Q_{n,\theta}\ra Q_{n,\theta}.$$ Then defining $\Pi_N:\Sc(\RR^d)\to\int_{\TT^d}^{\oplus}H_{N,\theta}\,d\theta$ by $$(\Pi_N(f))_{\theta}=\Pi_{N,\theta}(f),$$ we may show that $\Pi_N$ extends to a unitary operator $\Pi_N:L^2(\RR^d)\to\int_{\TT^d}^{\oplus}H_{N,\theta}\,d\theta$.

We can define pseudodifferential operators on $H_{N,\theta}$ analogously to (4.11) of the article, as $$\op_{N,\theta}(a):=\op_h(\tilde a)|_{H_{N,\theta}}.$$ which can be checked to map $H_{N,\theta}$ to itself. Furthermore one may show \cite[Section 2.2.3]{DyatlovJezequel2021} that $$\op_h(\tilde{a})=\Pi_N^*\left(\int_{\TT^{2d}}^{\oplus}\op_{N,\theta}(a)\,d\theta\right)\Pi_N.$$ In words, we have (up to unitary equivalence) expressed $\op_h(\tilde a)$ as a direct integral of the discrete operators $\op_{N,\theta}$. Moreover, the matrix elements $\la \op_{N,\theta}Q_{m,\theta},Q_{n,\theta}\ra$ vary smoothly with $\theta$; this may be checked by applying the formula $$U_wQ_{n,\theta}=e^{2\pi i\la\kappa,n-\theta_x\ra+\pi i N\la k,\kappa\ra}Q_{n,\theta-Nw}$$ in the cases $w_1=(k,\kappa), w_2=(k+\eps,\kappa+\eps')$ and invoking (\ref{eq:phasetranslation}) and (4.12) of the article.

Then a standard theorem on direct integrals \cite[Theorem XIII.83]{ReedSimonIV1978} gives that $$\esssup_{\theta\in\TT^{2d}}\|\op_{N,\theta}(a)\|_{H_{N,\theta}\to H_{N,\theta}}=\|\op_h(\tilde a)\|_{L^2(\RR^d)\to L^2(\RR^d)}
$$ and Equation (4.13) of the article follows from continuity in $\theta$.

\blfootnote{This work has been accepted by \textit{SIAM Multiscale Modeling \& Simulation}.}

\blfootnote{{For references of no fast-forwarding theorem, see, e.g., \cite[Theorem 3]{BerryAhokasCleveEtAl2007}, \cite[Theorem 5]{Childs2010CMP}, \cite{Kothari2010}, \cite{ZlokapaSomma2024}.}}

\end{document}